\documentclass[useAMS,usenatbib,letterpaper]{mn2e}
\usepackage{graphicx, amsmath, amssymb}
\usepackage{journal_abbreviations, custom_commands}
\usepackage{lastpage}
\usepackage[totalwidth=500pt,totalheight=650pt,centering,layoutvoffset=-0.3cm]{geometry}

\def\fwa{0.33\textwidth}
\def\fwb{0.45\textwidth}

\title[Metals in the $z\approx3.5$ IGM]
{Observations of metals in the \textit{z}$\, \approx\,$3.5 intergalactic medium and comparison to the EAGLE simulations\thanks{
Based on observations made with ESO Telescopes at the Paranal Observatory under programme IDs  091.A-0833, 092.A-0011 and 093.A-0575.}}
\author[Turner et al.]{Monica L. Turner,$^{1,2}$\thanks{E-mail: turnerm@mit.edu}
Joop Schaye,$^{2}$
Robert A. Crain,$^{3}$\newauthor
Tom Theuns,$^{4}$
and Martin Wendt$^{5,6}$\\
$^{1}$MIT-Kavli Center for Astrophysics and Space Research, 
   Massachusetts Institute of Technology, \\
  \,  77 Massachusetts Ave., Cambridge, MA 02139, USA\\
$^{2}$Leiden Observatory, Leiden University, PO Box 9513, 2300 RA Leiden, The Netherlands\\
$^{3}$Astrophysics Research Institute, Liverpool John Moores University, 146 Brownlow Hill, Liverpool, L3 5RF, UK\\
$^{4}$Institute for Computational Cosmology, Department of Physics, University of Durham, South Road, Durham, DH1 3LE, UK\\
$^{5}$Leibniz-Institut f{\"u}r Astrophysik Potsdam (AIP), An der Sternwarte 16, D-14482 Potsdam, Germany\\
$^{6}$Institut f{\"u}r Physik und Astronomie, Universit{\"a}t Potsdam, 14476 Potsdam, Germany\\
}

\begin{document}	

\date{\today}

\pagerange{\pageref{firstpage}--\pageref{LastPage}} \pubyear{2016}

\maketitle

\label{firstpage}

\begin{abstract}
We study the $z\approx3.5$ intergalactic medium (IGM) by comparing new,
high-quality absorption spectra of eight QSOs with  $\langle z_{\text{QSO}}\rangle=3.75$, to 
virtual observations of the EAGLE cosmological hydrodynamical simulations. 
We employ the pixel optical depth method 
and uncover strong correlations between various combinations of 
\hone, \cthree, \cfour, \sithree, \sifour, and \osix. 
We find good agreement between many of the simulated and observed
correlations, including $\tau_{\osixm}(\tau_{\honem})$.
However, the observed median optical
depths for the $\tau_{\cfourm}(\tau_{\honem})$ 
and $\tau_{\sifourm}(\tau_{\honem})$ relations
are higher than those measured from the mock spectra.  The discrepancy 
increases from up to $\approx0.1$~dex at  $\tau_{\honem}=1$ to 
$\approx1$~dex at $\tau_{\honem}=10^2$, 
where we are likely probing dense regions at small galactocentric 
distances.
As possible solutions, we invoke (a) models of ionizing radiation
softened above 4~Ryd to account for delayed completion 
of \hetwo\ reionization; (b) simulations run at higher resolution; (c) the inclusion of additional 
line broadening due to unresolved turbulence; and (d) increased
elemental abundances; however, none of these factors 
can fully explain the observed differences. Enhanced photoionization of
\hone\ by local sources, which was not modelled, could offer a solution.
However, the much better agreement with the observed
\osix(\hone) relation, which we find probes a hot and likely collisionally ionized
gas phase, indicates that the simulations are not in tension with the 
hot phase of the IGM, and suggests that the simulated outflows may entrain insufficient
cool gas.
\end{abstract}



\section{Introduction}
\label{sec:intro}

It is now well established that the high redshift
intergalactic medium (IGM) is enriched with heavy metals
to metallicites of $10^{-3}$ to $10^{-2}$~${\rm Z}_{\odot}$
\citep[e.g.,][]{cowie95, schaye03, simcoe04, aguirre08}.
While metals only constitute a fraction of the total baryon 
budget, they play an integral role in our understanding
of galaxy formation and evolution by providing a fossil record 
of star formation, and by impacting upon cooling-rates which can 
alter structure on many scales \citep[e.g.,][]{haas13a}. 

Metals are synthesized in and released 
from stars located in very overdense environments, 
therefore they need to travel large distances to reach the
diffuse IGM, and this transport is likely driven by 
feedback from star formation and active galactic nuclei (AGN).
Simulations have shown that metal pollution by galactic winds yields
reasonable enrichment statistics, without destroying the filamentary pattern 
that gives rise to the \hone\ \lya\ forest \citep{theuns02}. 
Although the need for inclusion of these processes
is clear, the mechanisms responsible
are not resolved, even in state-of-the-art cosmological simulations,
making their implementation uncertain. 
By comparing observed and theoretical
 metal-line absorption in the IGM, we may be able to constrain
 enrichment mechanisms such as outflows. 

Models and simulations of the IGM have been used
to make predictions about sources of metal pollution. 
\citet{booth12} established that the observations of \citet{schaye03} of
\cfour\ associated with weak \hone\ at $z\approx3$ 
can only be explained if the low-density IGM was enriched primarily by low-mass
galaxies ($M_{\text{halo}} \leq 10^{10}$~\msol) that drive
outflows to distances of $\sim10^2$~proper~kiloparsecs (pkpc),
and calculated that
$>10$\% of the simulated volume and $>50$\% of the baryonic mass 
in their successful model
was polluted by metals. 
The simulations studied by \citet{wiersma10}  indicate that 
at least half of the metals found in the $z=2$ IGM were ejected from 
galaxies at $z\geq3$, and that the haloes hosting these galaxies had masses less than
$M_{\text{halo}} = 10^{11}$~\msol.
This picture is consistent with that inferred from observations by
\citet{simcoe04}, who estimate that half of all baryons are enriched
to metallities $>10^{-3.5}$~Z$_{\odot}$ by $z\sim2.5$. 

Studies of the IGM using the direct detection of individual metal lines 
can typically probe only relatively
overdense gas, which constitutes a very small volume fraction of the 
Universe. In this work, we employ an approach known as the pixel 
optical depth method \citep{cowie98, ellison00, schaye00a, aguirre02, schaye03, turner14}, 
and provide a public version of the code at
\url{https://github.com/turnerm/podpy}.
This technique is a valuable tool for studying the IGM, as it enables
us to detect metals statistically even in low-density gas.
At the redshifts studied in this work, direct detection of metal-line absorption 
in regions of the spectrum contaminated by \hone\ is nearly
impossible due to the density of the forest of absorption features. By using 
the pixel optical depth method, we can correct for contamination and derive
statistical properties of the absorption by metals in this region. 
In this work we take advantage of the fact that this technique is
fast and objective, and can be applied uniformly to both observations
and simulations.

Our observational sample consists of new spectra of eight $3.62\leq z \leq 3.92$ QSOs
with uniform coverage and high signal-to-noise (S/N). We compare the results to the
Evolution and Assembly of Galaxies and their Environments (EAGLE)
cosmological hydrodynamical simulations \citep{schaye15, crain15}.
The EAGLE simulations are ideal for studying metal-line absorption in the IGM, as they 
have been run at relatively high resolution in a cosmologically 
representative volume ($2\times1504^3$ particles in a 100~cMpc box). 
The fiducial EAGLE model is able to reproduce the present-day galaxy stellar mass function, galaxy
sizes and the Tully Fisher relation \citep{schaye15}, and has been found to match observations
of galaxy colours \citep{trayford15} and the evolution of galaxy stellar masses \citep{furlong15}
and sizes \citep{furlong16}.
Furthermore, the simulations are in good agreement with a number of relevant observables, 
including the properties of \hone\ absorption at $z\sim2$--$3$ \citep{rahmati15},
as well as the \osix\ and \cfour\  \citep{schaye15}
and the \hone\ \citep{crain16} column density distribution functions (CDDFs) at $z\sim0$ (although 
for the latter we note that a higher than fiducial resolution is needed to 
achieve agreement).

\citet{rahmati16} compared observed metal-line CDDFs
for various ions and redshifts with the predictions from EAGLE, finding generally 
good agreement. Their study also included CIV, which was compared to the observations of
\citet[$z=1.5\text{--}5.5$]{songaila05},
\citet[$z=1.5\text{--}4.0$]{dodorico10}, and
\citet[$z=1.6\text{--}4.4$]{boksenberg15}.
Although the redshift ranges of these surveys overlap with ours, they are 
much wider and hence likely affected by evolutionary trends. These observational 
studies each used different, subjective methods to decompose the absorption 
into Voigt profiles, which limits their utility for testing models. Indeed, 
the different studies do not agree with each other, making it relatively easy 
for a single model to agree with the data. Here we will instead employ a like-for-like 
comparison of new data with virtual EAGLE observations tailored to mimic the 
characteristics of our own quasar spectra and analysed using the same automated 
method. We will consider relations between the pixel optical depths of 
\hone, \cthree, \cfour, \sithree, \sifour\ and \osix.

This paper is structured as follows.
In \S~\ref{sec:method}, we describe the observations and simulations.
We also summarize the pixel optical depth method, 
and how it is applied.
The results are presented in \S~\ref{sec:results}, 
and we give a discussion and conclusions in \S~\ref{sec:discussion} 
and \ref{sec:conclusion}, respectively.
Throughout this work, we denote proper and comoving 
distances as pMpc and cMpc, respectively. 
Both simulations and observations use cosmological parameters
determined from the \textit{Planck} mission \citep{planck13}, i.e.
$H_{\rm 0}=67.8$~\kmps~Mpc$^{-1}$, $\Omega_{\rm m} = 0.307$, 
and $\Omega_{\Lambda} = 0.693$.


\section{Method}
\label{sec:method}

\subsection{Observations}

We analyze a sample of eight QSOs with $3.62\leq z_{\text{QSO}} \leq 3.922$. 
They were selected based on their redshift and the existence
of substantial, high S/N data taken with VLT/UVES. Initially,
there were already 76.0 hours of UVES data, excluding overheads, 
of the QSOs.
Follow-up observations to fill in the gaps and improve S/N were 
completed in 62.7 hours of on-source time in 
programmes 091.A-0833, 092.A-0011 and 093.A-0575 (P.I. Schaye).
We note that for Q1422$+$23, the gaps in the UVES data were filled 
using archival observations with Keck/HIRES of comparable S/N and resolution
(which is $\approx8.5$~\kmps). 
The properties of the QSOs and the S/N of the spectra 
are summarized in Table~\ref{tab:qso}. 

The reduction of the UVES data was performed using the \texttt{UVES\_headsort} and 
\texttt{UVES\_popler} software by
Michael T. Murphy, and binned to have a uniform velocity dispersion 
of 1.3~\kmps.
The HIRES data was reduced using T. Barlow's \texttt{MAKEE} package, and binned on to 2.8~\kmps\ pixels. 
The continuum fits for the spectra were performed by hand by M.\ Turner. 
Any DLAs or Lyman break regions (i.e., due to
strong absorbers in \hone) were masked out, with the exception of DLAs in the \lya\ forest,
which were unmasked when recovering the \hone\ to be used for 
subtraction of contaminating absorption by higher-order Lyman
series lines from \osix\ and \cthree\ optical depths.

To homogenize the continuum fitting errors, we implemented the automated
sigma-clipping procedure of \citet{schaye03} at wavelengths greater than
that of the QSO's \lya\ emission, which was applied to both the observed and simulated spectra.
The spectrum is divided into rest-frame bins of $\Delta \lambda = 20$~\AA, 
which have central wavelength $\lambda_i$ and median flux $\bar{f_i}$. 
A B-spline of order 3 is then interpolated through all $\bar{f_i}$ values, and any pixels with flux
$N_{\sigma}^{\rm cf}\times\sigma$ below the interpolated values are discarded, where $\sigma$ is 
the normalized noise array. 
We then recalculate $\bar{f_i}$ without the discarded pixels, 
and repeat the procedure until convergence is reached. 
We use $N_{\sigma}^{\rm cf}=2$, which has been shown to be optimal in the \cfour\ region
for spectra with a quality similar to ours,
as it induces errors that are smaller than the noise by at least an order of magnitude \citep{schaye03}.

\begin{table*}
\caption{Properties of the QSOs used in this work, and the median S/N in the \hone\ \lya\ and \cfour\ 
recovery regions. The columns list, from left to right,
name, right ascension, declination, redshift, magnitude and magnitude band (either $V$-band (V), $R$-band (R),
or photographic (P)) from \citet{veron10},  
median signal-to-noise in the \lya\ forest
region and \cfour\ region, respectively (see \S~\ref{sec:redshift} 
for the definition of these regions), 
the median optical depth of \hone\ in the \lya\ forest, the factor $C_{\rm UVB}$ used to scale the UVB in the simulations,
and the redshift midpoint of eq.~\ref{eq:zlim} 
that defines the \lya\ forest region.}
\begin{center}	
\begin{tabular}{l l l l l l r r r r r}
\hline
 Name &		R.A.		&	Dec		&	$z_\text{QSO}$	&	Mag	&	Band	&	S/N$_{\lyam}$	&	S/N$_{\cfourm}$	& $\log_{10}\tau^{\rm med}_{\honem}$ & $C_{\rm UVB}$	&	$z^{\rm mid}_{\lyam}$\\                            
\hline	\hline													
Q1422$+$23 	&	\tt{	14:24:38	}	&	\tt{	+22:56:01	}	&	3.620	&	15.84	&	V	&	87	&	82	&	$-0.605$	&	0.896	&	3.24\\
Q0055$-$269	&	\tt{	00:57:58	}	&	\tt{	-26:43:14	}	&	3.655	&	17.47	&	P	&	60	&	79	&	$-0.583$	&	1.123	&	3.27\\
Q1317$-$0507	&	\tt{	13:20:30	}	&	\tt{	-18:36:25	}	&	3.700	&	18.10	&	P	&	59	&	90	&	$-0.549$	&	1.009	&	3.31\\
Q1621$-$0042	&	\tt{	16:21:17	}	&	\tt{	-23:17:10	}	&	3.709	&	17.97	&	V	&	78	&	92	&	$-0.546$	&	0.807	&	3.32\\
QB2000$-$330	&	\tt{	20:03:24	}	&	\tt{	-32:51:44	}	&	3.773	&	17.30	&	R	&	105	&	83	&	$-0.501$	&	0.809	&	3.38\\
PKS1937$-$101	&	\tt{	19:39:57	}	&	\tt{	-13:57:19	}	&	3.787	&	17.00	&	R	&	96	&	64	&	$-0.494$	&	1.293	&	3.39\\
J0124$+$0044	&	\tt{	01:24:03	}	&	\tt{	+00:44:32	}	&	3.834	&	18.71	&	V	&	48	&	59	&	$-0.445$	&	0.898	&	3.43\\
BRI1108$-$07	&	\tt{	11:11:13	}	&	\tt{	-15:55:58	}	&	3.922	&	18.10	&	R	&	29	&	29	&	$-0.390$	&	0.808	&	3.51\\	
\hline
\end{tabular}	
\end{center}
\label{tab:qso}
\end{table*}

\subsection{Simulations}

We compare the observations to predictions from the EAGLE cosmological 
hydrodynamical simulations. EAGLE was run with a substantially modified version of the $N$-body TreePM 
smoothed particle hydrodynamics (SPH) code \texttt{GADGET}~3
(last described in \citealt{springel05a}). EAGLE uses the package of hydrodynamics updates
``Anarchy''  (Dalla Vecchia, in prep.; see Appendix~A1 of \citealt{schaye15}) which invokes the pressure-entropy formulation 
of SPH from \citet{hopkins13}, the time-step limiter from \citet{durier12},
the artificial viscosity switch from \citet{cullen10}, an artificial conduction 
switch close to that of \citet{price08}, and the $C^2$ \citet{wendland95} kernel.
The influence of the updates is explored by \citet{schaller15}. 
The fiducial EAGLE model 
is run in a 100~cMpc periodic box with $1504^3$ of both dark matter and baryonic particles, 
and is denoted Ref-L1001504.  To test convergence with resolution and box size,
runs varying the number of particles and box size were also conducted, and are
listed in Table~\ref{tab:sims}.

The stellar feedback in EAGLE is implemented as described by \citet{dallavecchia12}, 
where thermal energy is injected  
stochastically. While the temperature of heated particles is always increased
by $10^{7.5}$~K, the probability of heating varies with the local metallicity and density \citep{schaye15, crain15}. 
The simulations include thermal AGN feedback \citep{booth09}, 
also implemented stochastically \citep{schaye15}. 
Both stellar and AGN feedback have been calibrated such that the simulations
match the observed $z\sim0$ stellar mass function and galaxy--black hole mass relation,
and give sensible disc galaxy sizes. We note that of the two highest-resolution runs,
Ref-L025N072 has been realized with the same subgrid parameters used in the fiducial model,
while for the Recal-L025N072 the subgrid parameters were re-calibrated to better match
the observed galaxy stellar mass function. 

EAGLE also includes a subgrid model for photo-heating
and radiative cooling via eleven elements: hydrogen, helium, carbon, nitrogen, oxygen,
neon, magnesium, silicon, sulphur, calcium and iron \citep{wiersma09a},
assuming a \citet{haardt01} UV and X-ray background. 
Star formation is implemented with a gas metallicity-dependent density threshold \citep{schaye04}
as described by \citet{schaye08}, followed by stellar evolution 
and enrichment from \citet{wiersma09b}.
Finally, details of the subgrid model for black hole seeding and growth 
can be found in  \citet{springel05b, rosas13} and \citet{schaye15}.

For each of our eight observed QSOs in Table~\ref{tab:qso},
we synthesize 100 corresponding mock spectra using the
\texttt{SPECWIZARD} package by Schaye, Booth, and Theuns (implemented 
as in \citealt{schaye03}, see also  Appendix~A4 of \citealt{theuns98}). 
To create mock spectra that resemble the observed QSOs and whose absorption features 
span a large redshift range, we follow \citet{schaye03} and stitch together
the physical state of the gas intersecting uncorrelated sightlines from 
snapshots with different redshifts. The ionization balance of each gas particle
is estimated using interpolation tables generated from \texttt{Cloudy} \citep[][version 13.03]{ferland13}
assuming uniform illumination by an ultra-violet background (UVB). 

\begin{figure}
 \includegraphics[width=0.5\textwidth]{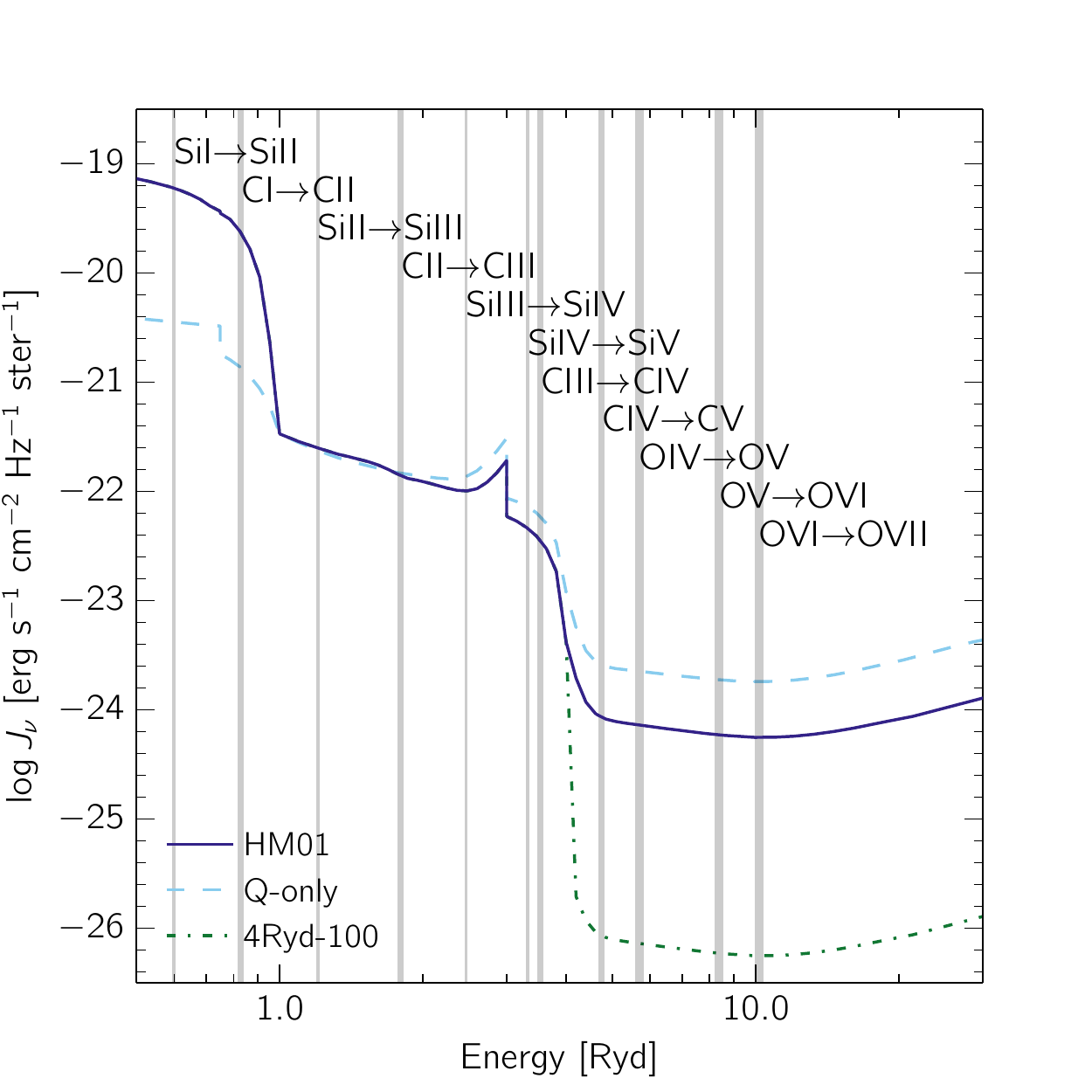}
 \caption{The intensity as a function of energy for the different UVB models.
  The different models are: HM01 QSO+galaxy \citep{haardt01}, which is our
 fiducial model;  Q-only, which is also by \citet{haardt01} but
 only considers an ionizing contribution from QSOs; and 
 4Ryd-100, the same as the fiducial model except that 	
 the intensity is reduced by a factor of 100 above 4~Ryd.
 The vertical light grey lines indicate the ionization energies
 of ions of interest, where increasing line thickness denotes silicon,
 carbon and oxygen ions, respectively. 
  All of the UVBs have been normalized
to have the same intensity as HM01 at 1~Ryd. 
}
\label{fig:uvb}
\end{figure}

We take the QSO+galaxy \citet{haardt01} UVB (denoted as ``HM01'') 
to be our fiducial model,\footnote{ Our choice of the HM01 UVB is primarily
to maintain consistency with the EAGLE simulations, in which the \citet{haardt01} UVB is
also used to calculate radiative cooling rates. 
While more recent
UVBs are available \citep[e.g.,][]{fauchergiguere09, haardt12}, they 
do not necessarily provide a better match to observations \citep[e.g.,][]{kollmeier14}.
}
and have plotted the intensity 
as a function of energy at $z=3.5$ in Fig.~\ref{fig:uvb}. 
We also consider
 the \citet{haardt01} background using quasars only (``Q-only''), which 
is much harder than the fiducial model above $\approx4$~Ryd.
Furthermore, to explore the possible effects of a 
delayed \hetwo\ reionization,  we consider a UVB that is 
significantly softer above 4~Ryd. To implement this,
we use the QSO+galaxy model and reduce the intensity above 4~Ryd
by a factor of 100, which we denote as ``4Ryd-100''. 

Self-shielding for \hone\ was included by modifying the ionization fraction
using the fitting functions of \citet{rahmati13a}. 
The normalization of the UVB is set such that the median recovered \hone\ \lya\ optical
depth of the simulated QSOs agrees with that of the observations at the same redshift. 
A unique value of $C_{\rm UVB}$ is determined for each
observed QSO and corresponding set of 100 mock spectra,
and the values are presented in Table~\ref{tab:qso}. 

In the EAGLE simulations, the dense particles 
that represent the multiphase interstellar
medium  (ISM, $n_{\rm H} > 0.1$~cm$^{-3}$)  are not allowed to cool 
below an effective equation of state (EoS),
and their temperature can be interpreted as a proxy for the pressure at which the warm 
and cool ISM phases equilibriate. 
We set their 
temperatures to $10^4$~K when generating
the mock spectra, although we note that due to the small cross-section 
of such dense absorbers the effect of including them is negligible. 

Each set of 100 mock spectra is synthesized to have redshifts 
identical to that of their corresponding observed QSO, and we consider absorption
ranging from $1.5 < z < z_{\text{QSO}}$ in every case. 
We include contributions from 31 \hone\ Lyman series transitions beginning with \lya,
and metal-line absorption from 
\ctwo, \cthree, \cfour, \fetwo, \nfive, \osix, \sitwo, \sithree, and 
\sifour\ (see Appendix~\ref{app:trans} for the rest wavelengths and oscillator strengths
 of these transitions). 
To match the spectral properties of the observations, the simulated spectra are convolved with a Gaussian 
with a FWHM of 6.6~\kmps, and resampled on to pixels of 1.3~\kmps. For each observed
QSO, we have measured the RMS noise in bins of 150~\AA\ in wavelength and 0.2 in  normalized flux.
We then use these measurements to add random Gaussian noise with the same variance to the simulations.

\begin{table*}
\caption{Characteristics of the EAGLE simulations. From left to right, the columns list the simulation
 name, box size, number of particles,
 initial baryonic particle mass, dark matter particle mass, comoving (Plummer-equivalent) gravitational softening, and
   maximum physical softening.}
   \begin{center}
\begin{tabular}{rrrccccl}
\hline
Simulation & $L$       & $N$ & $m_{\rm b}$ & $m_{\rm dm}$ & $\epsilon_{\rm com}$ & $\epsilon_{\rm prop}$ \\  
                & [cMpc] &       & [\msol]    & [\msol]        & [ckpc]          &    [pkpc]                       \\
\hline 
\hline
{Ref-L100N1504} &   100 & $2\times1504^3$ & $1.81 \times 10^6$ & $9.70 \times 10^6$ & 2.66 & 0.70  \\
{Ref-L050N0752} &    50  & $2\times752^3$ & $1.81 \times 10^6$ & $ 9.70 \times 10^6$ & 2.66 & 0.70   \\
{Ref-L025N0376} &    25  & $2\times376^3$ & $1.81 \times 10^6$ & $ 9.70 \times 10^6$ & 2.66 & 0.70  \\
{Ref-L025N0752} &    25  & $2\times752^3$ & $2.26 \times 10^5$ & $ 1.21 \times 10^6$ & 1.33 & 0.35 \\
{Recal-L025N0752} &    25  & $2\times752^3$ & $2.26 \times 10^5$ & $ 1.21 \times 10^6$ & 1.33 & 0.35 \\
\hline
\end{tabular}
\end{center}
\label{tab:sims}
\end{table*}

\subsection{Analyzed redshift range}
\label{sec:redshift}

The first step for the pixel optical depth 
recovery involves choosing optimal 
redshift limits. The fiducial redshift
range is selected to lie in the \lya\ forest, defined to be:
\begin{equation}
  (1 + \zqsom) \dfrac{\lambda_{\lybm} }{ \lambda_{\lyam}}- 1  \leq z \leq \zqsom - (1+\zqsom) \dfrac{3000\,\kmpsm}{c}
 \label{eq:zlim}
\end{equation}
where $\lambda_{\lyam}=1215.7$~\AA\ and $\lambda_{\lybm}=1025.7$~\AA\
are the \hone\ \lya\ and \lyb\ rest wavelengths, respectively. 
The lower limit was chosen to avoid the \lyb\ forest and corresponds 
to the \lyb\ transition at the 
redshift of the QSO, while the upper limit is 3000~\kmps\ bluewards of the QSO redshift
to avoid any proximity effects.

For \hone, \cfour\ ($\lrestm = [1548.2, 1550.8]$~\AA) and
\cthree\ ($\lrestm = 977.0$~\AA) we use the above
redshift limits. For the remaining ions, we make slight modifications, 
listed below,
in order to homogenize the contamination.
We use the notation $\lambda_{Z,k}$ to denote the rest wavelength of 
multiplet component $k$ of the ion $Z$. 
\begin{enumerate}
	\item \osix\ ($\lrestm=[1031.9, 1037.6]$~\AA): 
	We limit the recovery to where \osix\ overlaps with the \lyb\ forest and 
	place a cut-off at the \lya\ forest region, which leads to
      $z_{\rm max} = (1 + \zqsom) \lambda_{\honem,\lybm} / \lambda_{\osixm,2} - 1$
      \item \sithree\ ($\lrestm=1206.6$~\AA):
      We constrain the recovered optical depth region to not extend 
      outside of the \lya\ forest. For \sithree, which extends
      slightly bluewards into the \lyb\ forest, we set
      $z_{\rm min} =(1 + \zqsom) \lambda_{\lybm} / \lambda_{\sithreem}$.
      \item \sifour\ ($\lrestm = [1393.8, 1402.8]$~\AA):
      To avoid contamination by the \lya\ forest, we limit the blue end of the \sifour\ recovery 
      by setting 
      $z_{\rm min} = (1+\zqsom)\lambda_{\lyam}/\lambda_{\sifourm,1}-1$.
\end{enumerate}

\subsection{Pixel optical depth method}
\label{sec:pod}

We employ the pixel optical depth method, which we
use to study absorption on an individual pixel basis rather than
by fitting Voigt profiles to individual lines.
The goal is to obtain statistics on absorption by \hone\ and
various metal ions in the IGM, and on how their absorption relates to one another.
Our implementation is close to that of \citet{aguirre02}, but with 
the improvements of \citet{turner14}, and the effects of varying the 
chosen parameters can be found in both works. 
The exact methodology 
is described in full detail in Appendix~A of \citet{turner14},
and is summarized below. 
A working version of the code can be found at
\url{https://github.com/turnerm/podpy}.

After restricting the redshift range, the next step is to convert the flux of every
pixel of ion $Z$ and multiplet component $k$ to an optical depth
$\tau_{Z,k}(z) = -\ln(F)$, where $F(\lambda)$ is the normalised flux at
$\lambda = \lambda_{k}(1+z)$. Then, depending on the ion, corrections are made
for saturation or contamination, as described below. 
\begin{enumerate}
	\item For \hone\ \lya, while there is very little contamination in the \lya\ forest, the absorption in 
	many of the pixels will be saturated, and we use the higher order Lyman series
	transitions to correct for this. Specifically, if we consider a \lya\ pixel to be saturated, 
	we look to $N=16$ higher-order Lyman lines (beginning with \hone\ \lyb),
         and take the minimum optical depth, scaled to that of \lya, of all unsaturated pixels at the same redshift (if any). 
	If we are unable to correct the pixel due to saturation of the higher-order transitions,
        we set it to $10^4$ (so that these pixels can still be used for computing the median). Finally, we search for and discard
	any contaminated pixels, by checking that higher-order transitions
	do not have optical depth values significantly below what would be expected from the scaled \hone\ \lya\ optical depth.
	\item For \osix\ and \cthree, we can use the corrected \hone\ \lya\ optical depths
	to estimate and subtract contamination by \hone. 
	We do so beginning with \hone\ \lyb\ ($N=2$) and use higher-order Lyman series orders up to $N=5$.
	For saturated \osix\ and \cthree\ pixels, the optical depth is not well defined
	and therefore the above subtraction is not performed. Instead, we leave the pixel 
	uncorrected, unless the saturation can be attributed to \hone, in which case 
	the pixel is discarded. 
	\item \sifour\ and \osix\ are both closely-spaced doublets, and we can 
	use this fact to correct for contamination. To do so, we scale the optical depth
	of the weak component to match that of the strong component, and take the minimum of the
	two components modulo noise.
	We only take the scaled optical depth of the weaker component if it is significantly
	lower (when taking into account the noise array) than the stronger component.
	\item For \cfour, which is a strong transition redward of the \lya\ forest,
	the largest source of contamination is by its own doublet.
	To correct for this, we perform an iterative self-contamination correction.
	We first discard any pixels determined to be contaminated by other ions, by checking
        whether the optical depth of a pixel is too high
	to be explained by half of the associated stronger component combined with twice
	the associated weaker component. We then subtract the estimated contribution of the 
	weaker component from each pixel, iterating until convergence is reached.
\end{enumerate}

\begin{figure}
\begin{center}
 \includegraphics[width=\fwb]{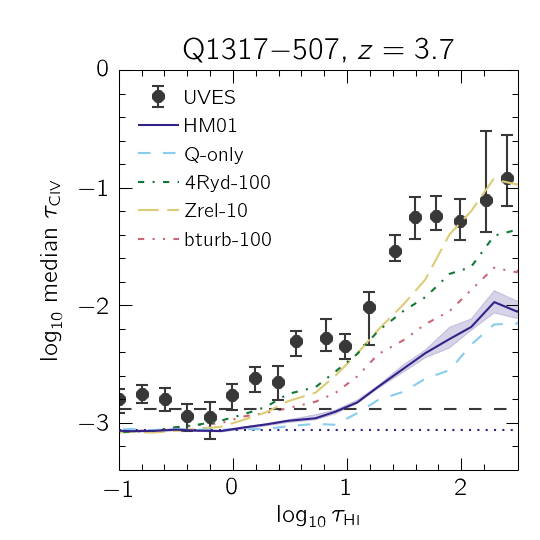}
\end{center}
 \caption{Example analysis   of the \cfour(\hone) relation for Q1317$-$507.
 The measured optical depths from the observations are shown by the points, and those recovered from the mock 
 spectra are indicated by the curves,
 with the different colours and line styles representing variations in the model. $1\sigma$ errors are indicated by the bars (shaded region)
 for the observations (fiducial simulation), and were determined by bootstrap resampling spectral chunks (mock spectra). 
 We indicate \taumin\ by the horizontal dashed and dotted lines for the observations and
 simulations, respectively.
 In Appendix~\ref{sec:single} we provide individual results for all eight QSOs, 
  and for all relations examined in this work. 
}
\label{fig:example}
\end{figure}

\subsection{Analysis}
\label{sec:pod_as}

For the analysis, we would like to see how the absorption from one ion
correlates with that from another. The procedures used here are also described in 
\S~3.4 and 4.2 of \citet{aguirre04}.
As an illustrative example, we will consider the ions
\cfour\ and \hone. For a single observed QSO, we use the recovered pixel
optical depths to construct a set of pixel pairs where each pair shares the same redshift. 
We then divide the ions into bins of $\log\tau_{\honem}$, and take the median
$\tau_{\honem}$ and $\tau_{\cfourm}$ in each bin	, to obtain
$\tau_{\cfourm}^{\text{med}}(\tau_{\honem})$, 
which from this point forward we will  denote as \cfour(\hone). 
The result of this procedure applied to
one of our QSOs is shown in Fig.~\ref{fig:example}, 
and we briefly describe the characteristics here.
We note that the results from individual QSOs for all 
relations examined here are given in Appendix~\ref{sec:single}.

We make note of two different regimes within the \cfour(\hone) relation. 
The first is on the right-hand side of Fig.~\ref{fig:example}, 
 where $\tau_{\honem}\gtrsim1$. Here, the median \cfour\ optical 
depth increases with \hone, which indicates that the pixels are probing gas enriched by \cfour.
The value of $\tau_{\cfourm}^{\text{med}}$ constrains the
number density ratio of \cfour\ to \hone.
Next, we turn to the region with $\tau_{\honem}\lesssim1$, where 
 $\tau_{\cfourm}^{\text{med}}$  is approximately constant.
 This behaviour arises because the median \cfour\ optical depths reach values
below the flat level \taumin, which is essentially a detection limit
set by noise, contamination, and/or continuum fitting errors.
An important caveat to keep in mind throughout this work is that 
the median recovered metal-line optical depth is not necessarily representative
of \textit{typical} intrinsic pixel optical depths for a given \hone\ bin.
In particular, as the metal-line optical
depths approach the flat level, it is likely that many individual pixels in a given
\hone\ bin have intrinsic metal optical depths at or below the flat level itself. In this case, the median
recovered metal optical depth will be  determined by the fraction of pixels
that have optical depths above the flat level. 

To construct the \cfour(\hone) relation for the 
observed spectra, \hone\ bins containing fewer than 25 pixels in total
are discarded. 
Furthermore, we divide each spectrum into chunks of 5~\AA\ (chosen to be much greater than the line widths),
and discard any bins containing fewer than 5 unique chunks. This is done to ensure that the 
median optical depths are obtained from more than just a few pixels in a very small spectral region.
To measure errors on $\tau_{\cfourm}^{\text{med}}$, 
we create new spectra by bootstrap
resampling the chunks 1000 times with replacement. We then measure 
\cfour(\hone)
for each bootstrap realization of the spectrum and take the error in each $\tau_{\honem}$ bin 
to be the $1\sigma$ confidence interval of all 
realizations. 

For the simulated spectra, we measure
\cfour(\hone)
for each mock spectrum, and require that each $\tau_{\honem}$ bin
have at least 5 pixels in total.
Next, we combine the results for all 100 mock spectra
associated with a single observed QSO by measuring
the median \cfour\ optical depth in each $\tau_{\honem}$ bin 
for all spectra, and we discard any bin containing contributions
from fewer than 5 spectra. 
Errors are calculated by bootstrap resampling the spectra 1000 times.

Next, we compute the flat levels \taumin\ by taking
the median of all pixels that have 
$\tau< \tau_c$, where we choose $\tau_c$ to be an optical depth below
which we do not find any correlations.  As in \citet{aguirre08}, we take $\tau_c=0.1$ when binning in \hone, 
and $0.01$ when binning in \cfour\ and \sifour. To estimate the error
on \taumin, for the observations we again divide the spectrum
into 5~\AA\ chunks, measure \taumin\ for 1000 bootstrap realizations, 
and take the $1\sigma$ confidence interval. For the simulations, we calculate
\taumin\ for each spectrum, and take the final value to be the median
value from all 100 spectra.

\begin{figure*}
\includegraphics[width=\fwa]{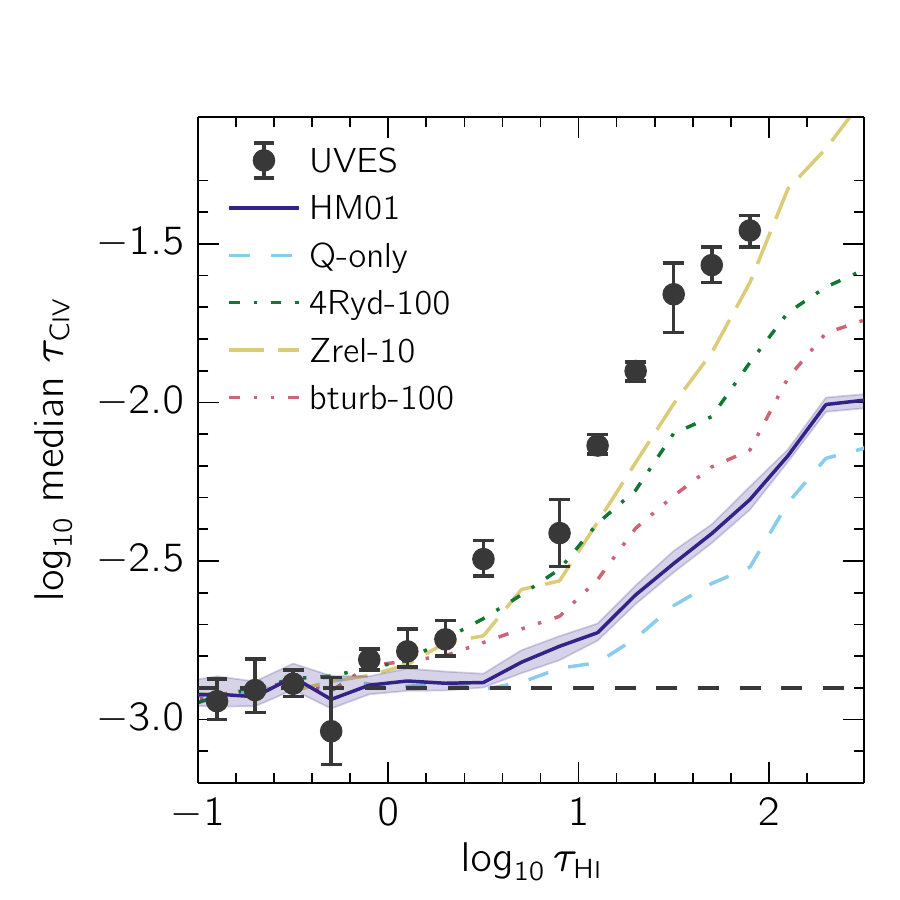}
\includegraphics[width=\fwa]{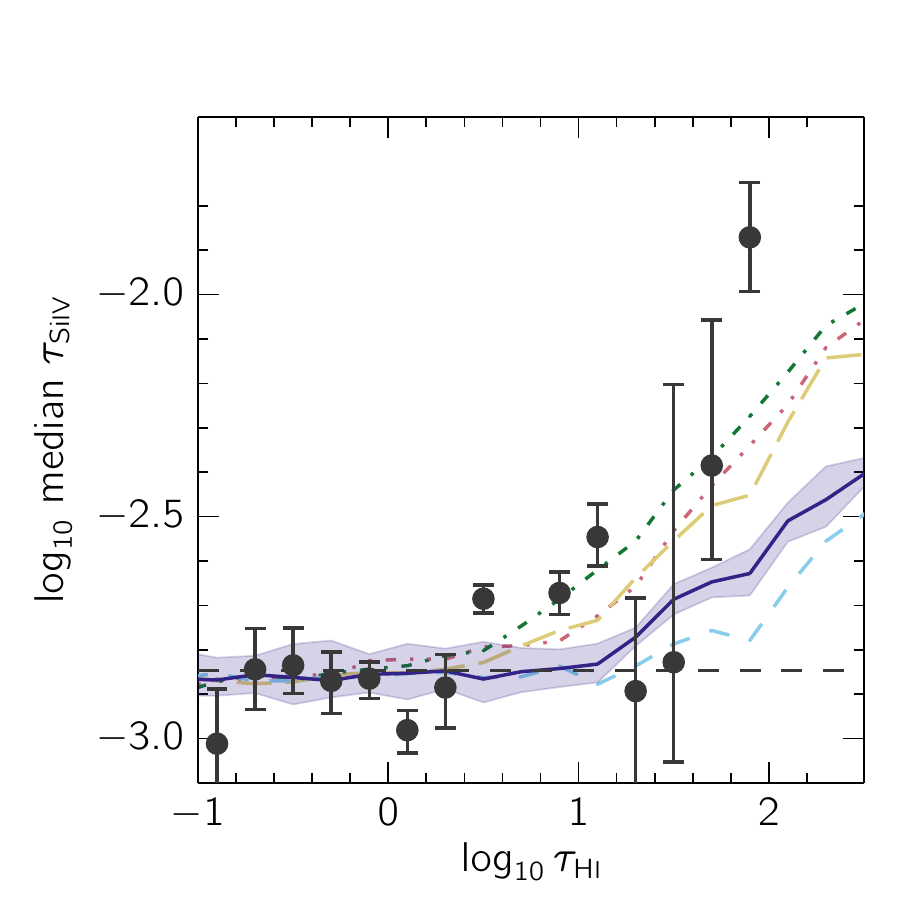}
\includegraphics[width=\fwa]{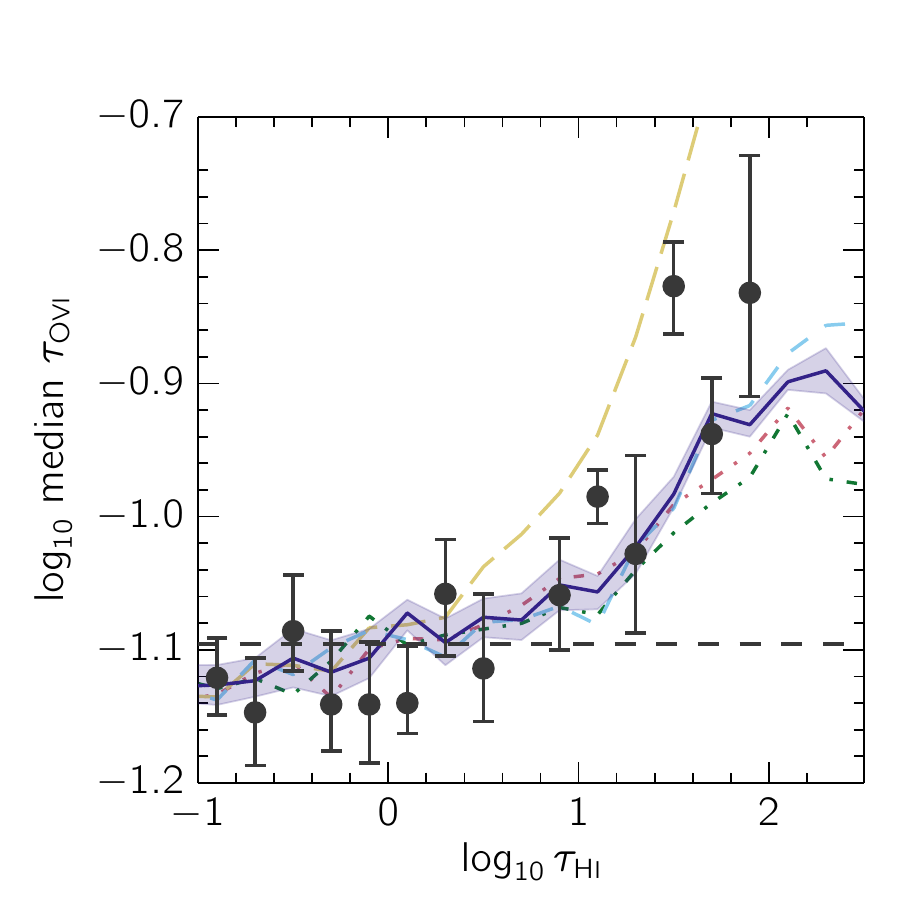}
 \caption{Median recovered pixel optical depths binned by \hone\ for \cfour\ (left), 
  \sifour\ (centre) and \osix\ (right). 
  Note that the dynamic range shown along the $y$-axis decreases strongly
  from left to right. 
  The data from eight QSOs have been
  combined, and the $1\sigma$ error bars are measured by bootstrap resampling the QSOs.
  The black circles show the data, while the curves denote the results from
  simulations, where different colours and line styles represent variations in the model and 
 we show the $1\sigma$ error region around the fiducial HM01 model. The flat level \taumin,
  which is the same for the observations and simulations by construction,
  is indicated by the dashed horizontal line. The data from the observations
  is provided in Table~\ref{tab:data_0}. 
 The 
  simulation run with the fiducial UVB systematically underpredicts the median
  \cfour\ and \sifour\ optical depths. The discrepancy is 
  lessened by invoking a softer UVB (4Ryd-100), higher metallicity (Zrel-10),
  or unresolved turbulent broadening (bturb-100), 
  although the metal-line optical depths associated with the strongest \hone\ absorption 
  are still underestimated by $\approx0.5$~dex. In contrast, the predicted 
  \osix(\hone) relation (right-hand panel) is insensitive
 to the UVB models, and in good agreement with the observations. 
}
\label{fig:hone}
\end{figure*}

Finally, below we outline the steps for combining the results from the different QSOs, 
which is applied to both the observed relations as well as their respective counterparts from the mocks.
Because our sample is uniform in terms of S/N, 
we simply combine the binned data points directly without subtracting \taumin.
 However, because the implementation of the noise, continuum fitting errors and contamination
in simulations is not completely accurate, the flat levels differ from the observations.
To account for this offset, we linearly add the difference between flat levels
($\tauminm^{\text{obs}} - \tauminm^{\text{sims}}$) to the median
optical depths in the simulations. We have verified that performing this step 
before the QSOs are combined does not modify the results. 
Next, to measure the combined median values, 
we perform $\chi^2$ fitting of a single value of $\tau_{\cfourm}^{\text{med}}$ to
all points in the bin, which is plotted against the central value of each \hone\ bin 
(in contrast to the results from individual QSOs, which are plotted against
the median of all \hone\ pixel optical depths in each bin).
We discard any data points that have contributions from fewer than four QSOs, 
and the $1\sigma$ errors are estimated by bootstrap resampling the QSOs. 
The combined results for \cfour(\hone) can be seen in the left-hand panel of Fig.~\ref{fig:hone}.


 \begin{table}
 \caption{Observational data shown in Fig.~\ref{fig:hone}. The left column 
   indicates the central $\log_{10}\tau_{\honem}$ value of each bin, and the next three columns list the $\log_{10}$ median
    recovered optical depths for the \cfour(\hone), \sifour(\hone), and \osix(\hone) relations, 
     respectively, along with the $1\sigma$ errors. We note that for the bin where 
     $\log_{10}\tau_{\honem}=0.70$, the metal optical depths are
     not defined because there were contributions from fewer that four of our eight quasars.}
 \begin{center}
\begin{tabular}{rrrr}
\hline$\log_{10}\tau_{\honem}$ & \cfour(\hone)  & \sifour(\hone)  & \osix(\hone) \\
\hline\hline$\tau_{\rm min}$ & $-2.90$ & $-2.85$ & $-1.10$\\
$-0.90$ & $-2.94_{- 0.06}^{+ 0.07}$ & $-3.01_{- 0.09}^{+ 0.12}$ & $-1.12_{- 0.03}^{+ 0.03}$\\
$-0.70$ & $-2.91_{- 0.07}^{+ 0.10}$ & $-2.84_{- 0.09}^{+ 0.09}$ & $-1.15_{- 0.04}^{+ 0.04}$\\
$-0.50$ & $-2.89_{- 0.04}^{+ 0.04}$ & $-2.83_{- 0.06}^{+ 0.08}$ & $-1.09_{- 0.03}^{+ 0.04}$\\
$-0.30$ & $-3.04_{- 0.10}^{+ 0.17}$ & $-2.87_{- 0.07}^{+ 0.07}$ & $-1.14_{- 0.04}^{+ 0.06}$\\
$-0.10$ & $-2.81_{- 0.03}^{+ 0.03}$ & $-2.87_{- 0.04}^{+ 0.04}$ & $-1.14_{- 0.04}^{+ 0.05}$\\
$ 0.10$ & $-2.79_{- 0.05}^{+ 0.07}$ & $-2.98_{- 0.05}^{+ 0.04}$ & $-1.14_{- 0.02}^{+ 0.04}$\\
$ 0.30$ & $-2.75_{- 0.05}^{+ 0.06}$ & $-2.88_{- 0.09}^{+ 0.07}$ & $-1.06_{- 0.05}^{+ 0.04}$\\
$ 0.50$ & $-2.49_{- 0.05}^{+ 0.06}$ & $-2.69_{- 0.03}^{+ 0.03}$ & $-1.11_{- 0.04}^{+ 0.06}$\\
$ 0.70$ & \ldots & \ldots & \ldots\\
$ 0.90$ & $-2.41_{- 0.10}^{+ 0.11}$ & $-2.67_{- 0.05}^{+ 0.05}$ & $-1.06_{- 0.04}^{+ 0.04}$\\
$ 1.10$ & $-2.14_{- 0.03}^{+ 0.04}$ & $-2.55_{- 0.07}^{+ 0.07}$ & $-0.98_{- 0.02}^{+ 0.02}$\\
$ 1.30$ & $-1.90_{- 0.03}^{+ 0.03}$ & $-2.89_{- 0.21}^{+ 0.21}$ & $-1.03_{- 0.06}^{+ 0.07}$\\
$ 1.50$ & $-1.66_{- 0.12}^{+ 0.10}$ & $-2.83_{- 0.22}^{+ 0.62}$ & $-0.83_{- 0.04}^{+ 0.03}$\\
$ 1.70$ & $-1.57_{- 0.05}^{+ 0.06}$ & $-2.38_{- 0.21}^{+ 0.33}$ & $-0.94_{- 0.04}^{+ 0.04}$\\
$ 1.90$ & $-1.46_{- 0.05}^{+ 0.05}$ & $-1.87_{- 0.12}^{+ 0.12}$ & $-0.83_{- 0.08}^{+ 0.10}$\\
\hline
\end{tabular}
\end{center}
\label{tab:data_0}
\end{table}

\section{Results}
\label{sec:results}

\subsection{$\tau_{Z}$ as a function of $\tau_{\rm HI}$}

We begin by examining median metal-line pixel optical 
depths as a function of \hone\ pixel optical depth in Fig.~\ref{fig:hone}, where we have plotted
\cfour(\hone), \sifour(\hone) and \osix(\hone) from left to right. 
The grey points with error bars represent the observations, 
while the curves show results from simulations, with different
colours indicating variations in the model.  The data from
the observations is presented in Table~\ref{tab:data_0}. 

The relations displayed in Fig.~\ref{fig:hone} depend on the following quantities 
in the simulations, using \cfour\ as an illustrative example: 
\begin{equation}
  \log_{10} \dfrac{\tau_{\cfourm}}{\tau_{\honem}}  = 
  \log_{10} \dfrac{(f\lambda)_{\cfourm}}{(f\lambda)_{\honem}}
   \dfrac{n_{\cfourm}}{n_{\rm C}}
   \dfrac{n_{\rm H}}{n_{\honem}} 
   + [\text{C}/\text{H}] 
    + (\text{C}/\text{H})_{\odot} ,
\label{eq:metallicity}
\end{equation}
where $f$ is the oscillator strength, $\lambda$ is the rest wavelength, and
$n$ is the number density.
While the oscillator strength and rest wavelength are fixed
empirical quantities, the element abundances are predicted, and 
the ionization fractions (i.e., the ratio of ionized to total number density)
are determined using the particle temperatures and densities in the same manner
as was done to compute the cooling rates used in the simulation. 

In the following analysis, we will consider the results in 
two different regimes, separated by $\tau_{\honem}\approx10$. The reasons for this are: 
(1) For $\tau_{\honem}\gtrsim10$ \hone\ pixel optical depths
will be highly saturated, and even though this is corrected for in our recovery procedure, the final values
still suffer from large uncertainties compared to their unsaturated counterparts.
(2) If the gas being probed is mainly in photoionization equilibrium,
which is a reasonable assumption for \cfour\ and \sifour\ \citep{schaye03, aguirre04}, then
\hone\ is considered a good tracer of the density \citep{schaye01}, even on an individual 
pixel basis \citep{aguirre02}. 
This means that higher \hone\ optical 
depths probe dense regions closer to galaxies,
rather than the diffuse IGM. 

\begin{figure}
\begin{center}
\includegraphics[width=\fwb]{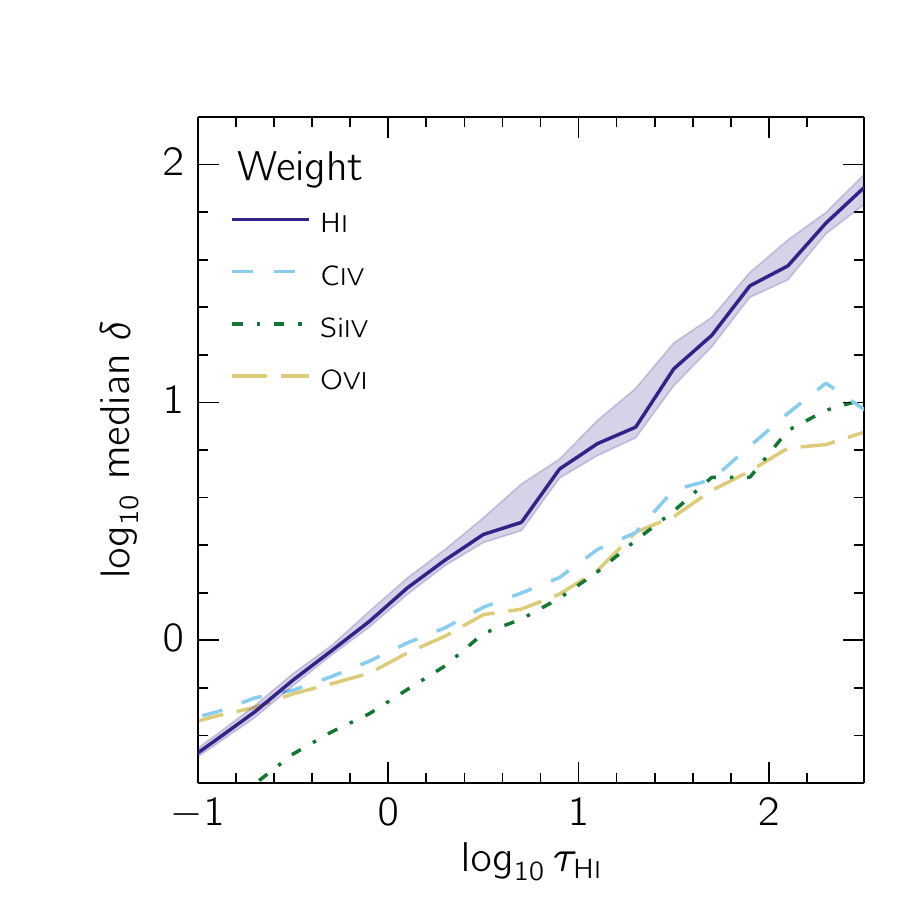}
\end{center}
 \caption{Median optical depth-weighted overdensity as a function of \hone\ optical depth
 for the simulated spectra. The different curves indicate the ion used for the 
 weighting. We have indicated the one $1\sigma$ error region around the \hone-weighted overdensity. 
 The parameters from an ordinary least-squares fits of the form 
 $\log_{10} \delta =A \times \log_{10} \tau_{\honem} +B$ to each of the relations
 are given in Table~\ref{tab:delta}. The metal-line absorption typically arises in lower-density 
 gas than the \hone\ absorption with the same redshift.
 }
\label{fig:delta}
\end{figure}

To touch on this point more quantitatively, we have calculated
the optical depth-weighted overdensities for 
our full sample of mock spectra.  Fig.~\ref{fig:delta} shows the median optical depth-weighted
overdensity as a function of \hone\ for
\hone, \cfour, \sifour, and \osix, and we find a strong correlation 
in every case. We have performed ordinary least-squares
fits to the curves using a power-law function:
\begin{equation}
	\log_{10} \delta =A \times \log_{10} \tau_{\honem} + B,
	\label{eq:delta_vs_hone}
\end{equation}
and give the resulting parameters in Table~\ref{tab:delta}. We note that our aforementioned division of
the \hone\ optical depths into two regimes at $\tau_{\honem}\approx10$ corresponds to an overdensity 
of $\approx8$ for the gas responsible for the \hone\ absorption, but that the overdensity of the metal 
ions at the same redshifts as the selected \hone\ pixels is typically a factor of a few lower. 
We also caution that the quantities presented here are pixel-weighted,
and may therefore be biased to high temperatures since higher temperature lines will be more broadened 
and therefore cover more pixels. This may also bias the densities to lower values, as intergalactic gas with $T\gtrsim10^5$~K
tends to correspond to lower densities than that found at lower temperatures. 
\citep[e.g., Fig.~1 from][]{vandevoort11a}. 

The fitted values for \hone\ can be compared to eq.~8 of \citet{rakic12}, who used 
the relations of \citet{schaye01} to obtain an expression for overdensity as a
function of \hone\ \lya\ line centre optical depth $\tau_{0,\lyam}$, under the 
assumption that the absorbers have 
sizes of the order of the local Jeans length. Modifying their eq.~8
using the parameters from this work, we obtain:
\begin{equation}
\begin{split}
	\delta \approx& \, 0.74\, \tau_{0,\lyam}^{2/3}\
	\left(\dfrac{\Gamma_{12}}{0.85\times10^{12}\text{ s}^{-1}} \right)^{2/3} 
	\left( \dfrac{1+z}{4.36}\right)^{-3} \\
	\times& \left(\dfrac{T}{2\times10^4\text{ K}} \right)^{0.17} 
	\left( \dfrac{f_{\rm g}}{0.15} \right)^{-1/3}
	\left( \dfrac{b}{20 \text{ \kmps}} \right)^{2/3}.
\end{split}
\label{eq:rakic_eq8}
\end{equation}
To determine the redshift $z$, for each QSO we considered the 
redshift at the centre of the \lya\ forest as defined by eq.~\ref{eq:zlim}, 
and took the mean of all of the values. To obtain the photoionization rate
$\Gamma_{12}$, we multiplied the photoionization rate from \citet{haardt01} at 
each of the above redshifts
by the scale factor that was used to bring the median recovered \hone\ optical depth of 
the mock spectra into agreement with that of the corresponding observed spectrum, and took the mean for 
all set of mocks. As in \citet{rakic12}, we chose a temperature typical of 
a moderately-dense region in  IGM  \citep[e.g.,][]{schaye00b, lidz10, becker11, rudie12b},
and assumed that the gas fraction $f_{\rm g}$ corresponds to the universal value of $\Omega_{\rm b}/\Omega_{\rm m}$.
Finally, we have taken $b$ to be $20$~\kmps, which is similar to
values measured by \citet{rudie12} for $z\approx2.5$. 
Putting eq.~\ref{eq:rakic_eq8} into the form of eq.~\ref{eq:delta_vs_hone}, we obtain 
$A=0.67$ and $B=-0.13$, compared with $A=0.66\pm0.01$ and $B=0.16\pm0.01$ measured from our simulations.
Hence, the slope $A$ is in excellent agreement with the theoretical scaling relation implied by the model
of \citet{schaye01}. The normalization agrees to within a factor of two, which we consider
good agreement given the uncertainties in the fiducial parameter values. 

\begin{table}
 \caption{Results from fitting a power-law $\log_{10} \delta =A \times \log_{10} \tau_{\honem} +B$ to
	the relations in Fig.~\ref{fig:delta}, using ordinary least-squares.}
 \begin{center}
\begin{tabular}{lrr}
\hline
Ion & $A$ & $B$ \\	
\hline \hline 
  \hone & $ 0.66\pm 0.01$ & $ 0.16\pm 0.01$ \\
 \cfour & $ 0.37\pm 0.02$ & $-0.02\pm 0.01$ \\
\sifour & $ 0.52\pm 0.01$ & $-0.24\pm 0.01$ \\
  \osix & $ 0.36\pm 0.02$ & $-0.05\pm 0.01$ \\
  \hline
\end{tabular}
\end{center}
\label{tab:delta}
\end{table}

With the above in mind, we can interpret the results of  Fig.~\ref{fig:hone}. 
Focusing first on the left-hand panel, we find that at fixed \hone,  the observed 
median \cfour\ optical depths are significantly higher than in the fiducial HM01 model.
The discrepancy increases from $\approx0.1$~dex at $\tau_{\honem}=1$ to $\approx0.5$~dex
at $\tau_{\honem}=10$ and $\approx1$~dex at $\tau_{\honem}=10^2$.
This suggests that at a given
gas overdensity, there is less \cfour\ in the simulations by $\approx0.5$~dex ($\approx1.0$~dex)
for $\tau_{\honem}\lesssim10$ ($\tau_{\honem}\gtrsim10$). 

Turning to the different UVB models, while the harder Q-only background
provides a poorer match to the observations, 
4Ryd-100 fares much better. Although this model still falls short of the 
observed $\tau_{\cfourm}^{\text{med}}$ by about $\approx0.5$~dex
in the highest $\tau_{\honem}$ bin, the softest background is nearly
fully consistent with the observations for $\tau_{\honem}\lesssim10$.

From eq.~\ref{eq:metallicity},
it is apparent that an increase in [C/H] will lead to higher \cfour\ optical depths
at fixed \hone. Therefore, we have run \texttt{SPECWIZARD} with the 
elemental abundances scaled linearly by a factor of ten, denoted as ``Zrel-10''. 
Such a modification can be partially motivated by the fact that we expect some uncertainty 
in the nucleosynthetic yields, of about a factor of two. We have chosen a factor much larger than this
since we find that increasing the metallicities does not scale the median recovered optical depth by the same factor 
(contrary to what one might expect from eq.~\ref{eq:metallicity}). 
This is because many pixels contributing to the median optical depth are noise dominated
(particularly at low \hone), and some fraction of these pixels have a true optical depth of zero,
so no matter how much the metallicity is increased, the optical depth will never change and exceed the noise. 

Although Zrel-10 provides much better agreement between the model and data, even 
with such an extreme choice of multiplicative factor it cannot fully account 
for the discrepancy between the simulations and data.
This suggests that in the simulation too many \hone\ clouds have negligible metallicity. 
To check this, we have calculated the mass
and volume filling factors of the metals using eq.~1 from
\citet{booth12}\footnote{We compute the mass filling factors 
using SPH smoothed metallicities, but to avoid smoothing twice, 
we compute the volume filling fractions using the particle 
metallicities (see \citealt{wiersma09b} for a discussion on the use of 
SPH-smoothed versus particle metallicities).} 
The authors determined that $>10$\% of the volume and $>50$\% of 
the mass need to be enriched to metallicities $Z> 10^{-3}$~Z$_\odot$ to 
achieve agreement with observations of \cfour\ in the low-density IGM 
at $z\approx3$. For the fiducial L100N1504 box at $z=3.5$ and $Z>0$ we 
find volume and mass filling factors of 42\% and 68\%, but for 
$Z>10^{-3}$~Z$_\odot$ these are reduced to 10\% and 19\%, respectively. 
This suggests that while the fractions of the volume and mass with 
non-zero metallicity may be sufficiently high, the metallicities 
in the photo-ionised IGM are typically far too low.

Next, we describe another modification to our fiducial model denoted as 
``bturb-100'', for which we have considered an unresolved
turbulent broadening term in addition to the
usual thermal broadening. Specifically, we add
$b_{\rm turb} = 100$~\kmps\ in quadrature
to the already included thermal broadening 
$b_{\rm therm} (T, m)$ which is calculated for every ion
at each spectral pixel, and 
depends on the local temperature $T$ and inversely on atomic mass $m$. 
Because of this inverse dependence on atomic mass,
metal ions will be much more strongly affected by the 
inclusion of turbulent broadening than \hone.\footnote{
We initially added the thermal broadening term to both \hone\ and metal ions. However,
this led to unphysical values for the flat level of metal ions recovered from
regions bluewards of the \lya\ forest, due to the extreme abundance of \hone\ lines. 
Thus, for the model presented in this work we have added turbulent broadening only to 
the metals and not to \hone, which 
allows us to examine the effect of broadening on \osix, \cthree\ and \sithree\ more clearly.
This is likely also more physical, as the metal-bearing gas may well 
be more turbulent than the gas that dominates the neutral hydrogen 
absorption \citep[e.g.,][]{theuns02}. 
We have confirmed that both methods produce equivalent results
for the \cfour(\hone) and \sifour(\hone) relations, likely because 
the metal-line optical depth signal is mostly associated with the strongest and most clustered
hydrogen systems that are not significantly affected by an increased $b$-parameter.
}
Indeed, we find that bturb-100 provides a somewhat better
match to the observed \cfour(\hone) relation. 
However, we note that our chosen broadening value $100$~\kmps\ should be considered 
a very conservative upper limit, as individual 
\cfour\ and \sifour\ components are not usually detected with such large $b$-parameters.

In the centre panel of Fig.~\ref{fig:hone}, we show \sifour(\hone), and find
results that are similar to those for \cfour(\hone). 
For $\tau_{\honem}\gtrsim10$ the \sifour\ optical 
depths are underestimated by the fiducial HM01 model by a factor
ranging from $\approx0.2$~dex at $\tau_{\honem}=10$
up to $\approx0.8$~dex at $\tau_{\honem}=10^2$. 
Invoking 4Ryd-100 leads to near agreement
for all but the highest \hone\ optical depth, while Zrel-10 and bturb-100
also fare markedly better than the fiducial model.

We now consider the \osix(\hone) relation in the right-hand panel of Fig.~\ref{fig:hone}. 
While \cfour\ and \sifour\ are expected to mainly probe
cool photoionized gas ($T\sim10^4$~K), \osix\ reaches its peak ionization fraction of 0.2 at
$T=3\times10^5$~K, which is close to the temperatures expected of shocks
associated with accretion events or winds.
Simulations predict that \osix\ around galaxies
is primarily collisionally ionized \citep[e.g.][]{tepper-garcia11, stinson12, ford13, shen13}.
Applying ionization modelling to observations also provides
evidence that \osix\ near moderate to strong \hone\ preferentially probes this 
hot gas phase \citep[e.g,][]{simcoe04, aguirre08, danforth08, savage14, turner15}. 

Indeed, the results from the right-hand panel of Fig.~\ref{fig:hone} differ considerably 
from the previous two relations. Firstly, the simulation realized with the
fiducial model is almost fully consistent with the observations, with any discrepant 
points offset by a maximum of 0.2~dex (note the smaller dynamic range
of the y-axis compared to the previous panels). 
While 
the alternate UVBs (Q-only and 4Ryd-100) have slightly lower $\tau_{\osixm}$ than the fiducial case, 
overall we do not find significant differences between these models. 
This suggests that in EAGLE the \osix(\hone) relation may be probing a 
primarily collisionally ionized gas phase, for which variations in the ionization background
do not have a significant impact on the results.  
Furthermore, the fact that the median \osix\ optical depth
appears to be largely insensitive to the addition of turbulent broadening 
could indicate that \osix\ is already significantly thermally broadened. 
We note that if the pixel optical
depths do not originate predominantly from photoionized gas, then $\tau_{\honem}$ can no
longer be used as a measure of the density.

\begin{figure*}
 \includegraphics[width=\fwb]{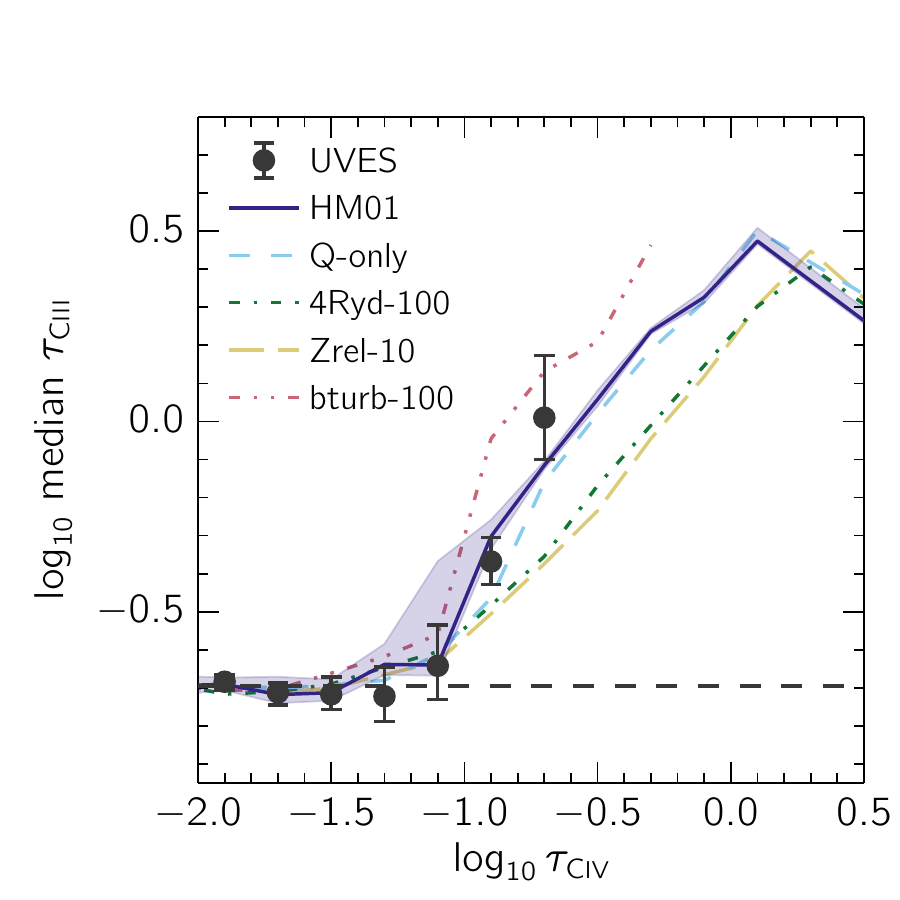}
  \includegraphics[width=\fwb]{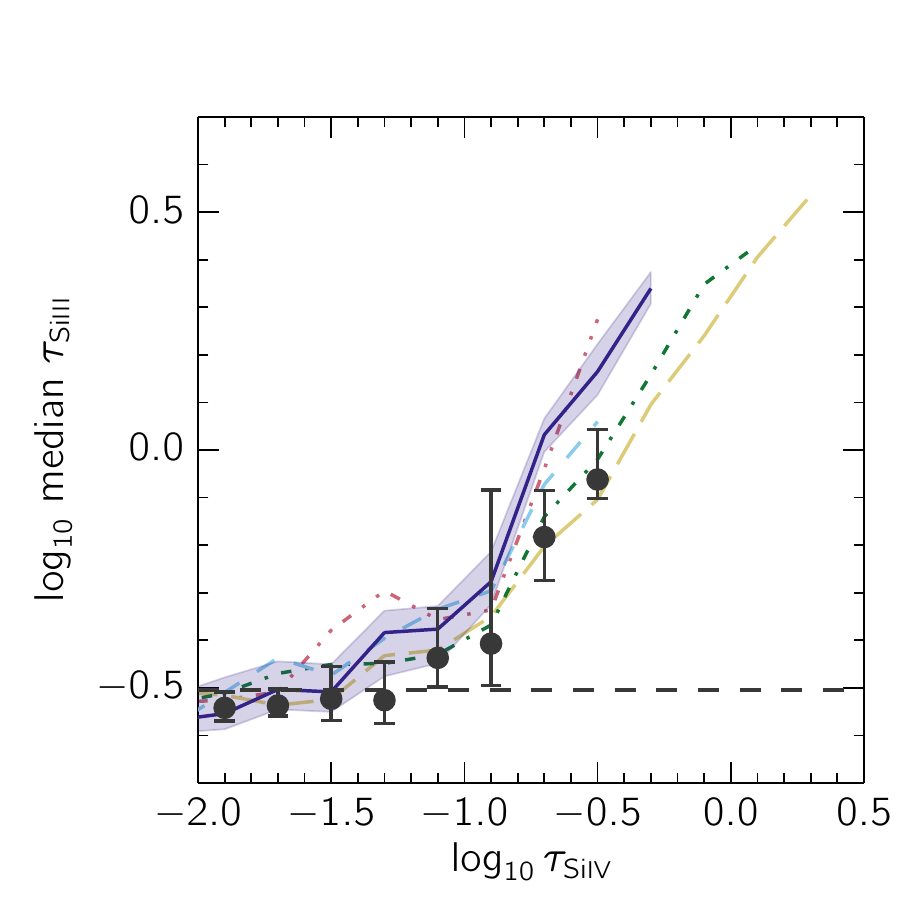}\\
  \caption{The same as Fig.~\ref{fig:hone}, but showing \cthree(\cfour)
 and \sithree(\sifour), and with the data from observations presented in 
   Table~\ref{tab:data_1}. Unlike for relations binned by \hone,  
   different ionization states of the same element are not sensitive to 
   the metallicity of the gas. We find that for \cthree(\cfour), the simulations and data
  are in good agreement for the fiducial ionizing background, and the observations
  particularly disfavour 4Ryd-100 and Zrel-10. The \sithree(\sifour) relation is somewhat
  less constraining, and while the median \sithree\ optical depths from HM01 model 
  are slightly above the observed values, the discrepancy is no more than 0.1~dex and only
  seen in the highest \sifour\ bins. 
  This indicates that 
  the temperature and density of the gas probed by pixels with detected \cfour\ and and \sifour\ is 
  well captured by the simulations, without needing to invoke modifications to the model.}
\label{fig:temp}
\end{figure*}

 \begin{table*}
 \caption{Observational data from Figs.~\ref{fig:temp} and \ref{fig:rel}. 
 The format is the same as Table~\ref{tab:data_0}, but here we present
  relations binned by either \cfour\ or \sifour\ optical depths. The left column
  indicates the central value of the \cfour\ or \sifour\ bin, and the 
   subsequent columns list the median recovered optical depths for the relation
   denoted in the top row. }
\begin{tabular}{rrrrrr}
\hline$\log_{10}\tau_{\rm bin}$ & \cthree(\cfour)  & \sithree(\sifour)  & \sifour(\cfour)  & \osix(\cfour)  & \osix(\sifour) \\
\hline\hline 
$ \tau_{\rm min}$ & $-0.70$ & $-0.50$ & $-2.78$ & $-1.06$ & $-1.10$\\
$-1.90$ & $-0.68_{- 0.02}^{+ 0.02}$ & $-0.54_{- 0.03}^{+ 0.03}$ & $-2.84_{- 0.03}^{+ 0.03}$ & $-1.17_{- 0.05}^{+ 0.08}$ & $-1.12_{- 0.04}^{+ 0.03}$\\
$-1.70$ & $-0.71_{- 0.03}^{+ 0.03}$ & $-0.54_{- 0.02}^{+ 0.04}$ & $-2.82_{- 0.06}^{+ 0.05}$ & $-1.07_{- 0.03}^{+ 0.03}$ & $-1.10_{- 0.02}^{+ 0.03}$\\
$-1.50$ & $-0.72_{- 0.04}^{+ 0.04}$ & $-0.52_{- 0.05}^{+ 0.07}$ & $-2.64_{- 0.04}^{+ 0.05}$ & $-1.02_{- 0.02}^{+ 0.02}$ & $-1.08_{- 0.04}^{+ 0.04}$\\
$-1.30$ & $-0.72_{- 0.07}^{+ 0.08}$ & $-0.53_{- 0.05}^{+ 0.08}$ & $-2.45_{- 0.05}^{+ 0.05}$ & $-1.05_{- 0.04}^{+ 0.04}$ & $-0.99_{- 0.05}^{+ 0.03}$\\
$-1.10$ & $-0.64_{- 0.09}^{+ 0.11}$ & $-0.44_{- 0.06}^{+ 0.10}$ & $-2.21_{- 0.05}^{+ 0.05}$ & $-0.95_{- 0.02}^{+ 0.02}$ & $-0.99_{- 0.04}^{+ 0.05}$\\
$-0.90$ & $-0.37_{- 0.06}^{+ 0.06}$ & $-0.41_{- 0.09}^{+ 0.32}$ & $-1.98_{- 0.05}^{+ 0.07}$ & $-1.00_{- 0.09}^{+ 0.15}$ & $-0.86_{- 0.03}^{+ 0.04}$\\
$-0.70$ & $ 0.01_{- 0.11}^{+ 0.16}$ & $-0.18_{- 0.09}^{+ 0.10}$ & $-1.57_{- 0.10}^{+ 0.18}$ & $-0.66_{- 0.07}^{+ 0.08}$ & $-0.81_{- 0.09}^{+ 0.29}$\\
$-0.50$ & \ldots & $-0.06_{- 0.04}^{+ 0.10}$ & $-1.07_{- 0.20}^{+ 0.19}$ & $-0.43_{- 0.04}^{+ 0.03}$ & $-0.90_{- 0.15}^{+ 0.54}$\\
$-0.30$ & \ldots & \ldots & $-1.87_{- 0.17}^{+ 0.36}$ & $-0.82_{- 0.09}^{+ 0.31}$ & \ldots\\
$-0.10$ & \ldots & \ldots & $-0.82_{- 0.01}^{+ 0.34}$ & $-0.40_{- 0.00}^{+ 0.04}$ & \ldots\\
\hline
\end{tabular}
\label{tab:data_1}
\end{table*}

\subsection{$\tau_Z$ as a function of $\tau_{\rm CIV}$ and $\tau_{\rm SiIV}$} 

While metal ions as a function of $\tau_{\honem}$ can probe the metallicity-density relation, 
examining different ionization states of a single element can provide insight into
the physical properties of the gas, because the ionization fractions that
set the relative optical depths will only depend on the temperature, the density, 
and the UV radiation field (and not on the metallicity, but see below). These optical depth ratios
have previously been used to establish that the gas probed by \cfour\ and \sifour\ is
consistent with being in photoionization equilibrium \citep{schaye03, aguirre04}. 

Fig.~\ref{fig:temp} examines \cthree(\cfour) and \sithree(\sifour), and the 
 observational data is provided in Table~\ref{tab:data_1}. 
Looking first at \cthree(\cfour), we find that HM01 is consistent 
with all of the \cfour\ bins. Both the bturb-100 and Q-only models 
also agree with the data, which is notable in particular for Q-only as it is the most disfavoured by the 
\cfour(\hone) relation. 
Finally, we find that the 4Ryd-100 and Zrel-10 models fare 
particularly poorly in this relation, and produce median \cthree\ optical depths lower than
the observations by up to 0.4~dex. In particular, for Zrel-10 such a discrepancy
may seem surprising, since changing the carbon abundance should not affect 
the amount of one ionization state against another. However, the reason this occurs
is because \cfour\ increases more than \cthree, which is
due to the fact that many more \cthree\ pixels are contaminated and hence do not change. 

Next, we find the \sithree(\sifour) relation
to be somewhat less constraining. While the fiducial HM01 model demonstrates one of the largest
discrepancies with the data,
the difference is not more than $\approx0.1$~dex when the errors are considered, 
and is only seen in the highest \sifour\ bins.
Thus, we find good agreement between the data and the fiducial model for 
both relations, which suggests that the temperature and density of the gas probed by 
\cfour\ and \sifour\ pixels is consistent between the observations and simulations.

We note that the above result is not in tension with the results shown in Fig.~\ref{fig:hone}. Namely,
while Fig.~\ref{fig:hone} indicates that there is a lack of \cfour\ and \sifour\ in the simulations,
Fig.~\ref{fig:temp} demonstrates that the \sifour\ and \cfour\ that we do find in the mock spectra,
regardless of the amount,
likely resides in gas with similar temperature and density as in the observations.

\begin{figure*}
 \includegraphics[width=\fwa]{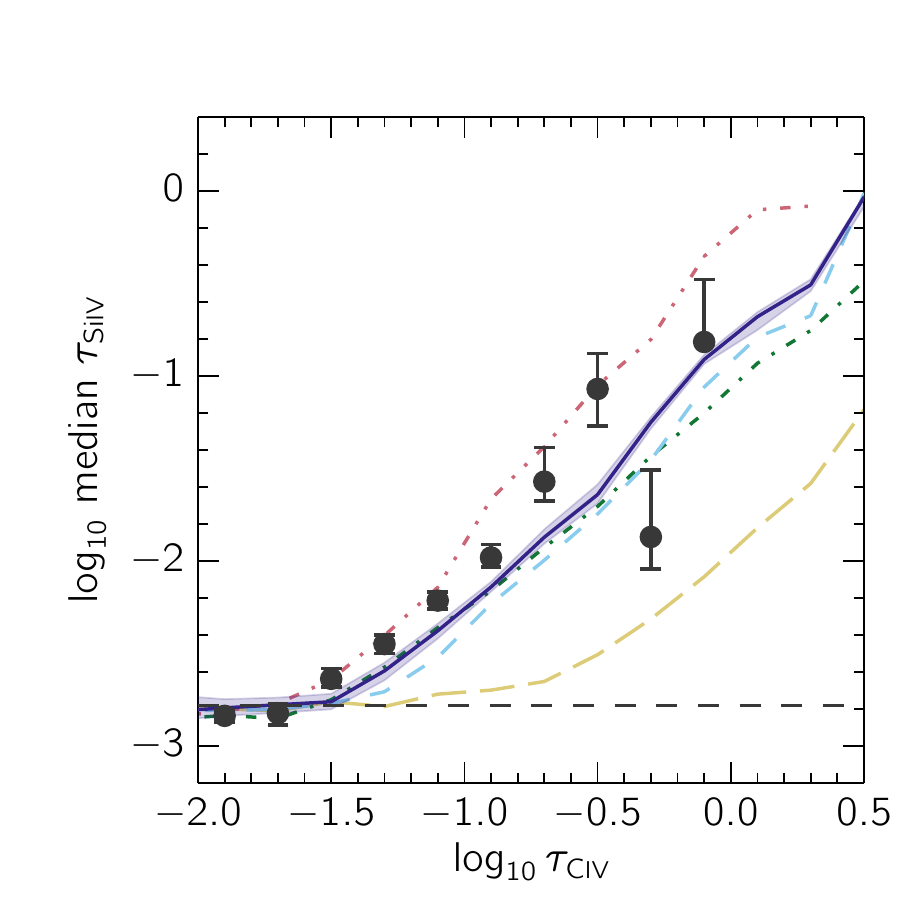}
 \includegraphics[width=\fwa]{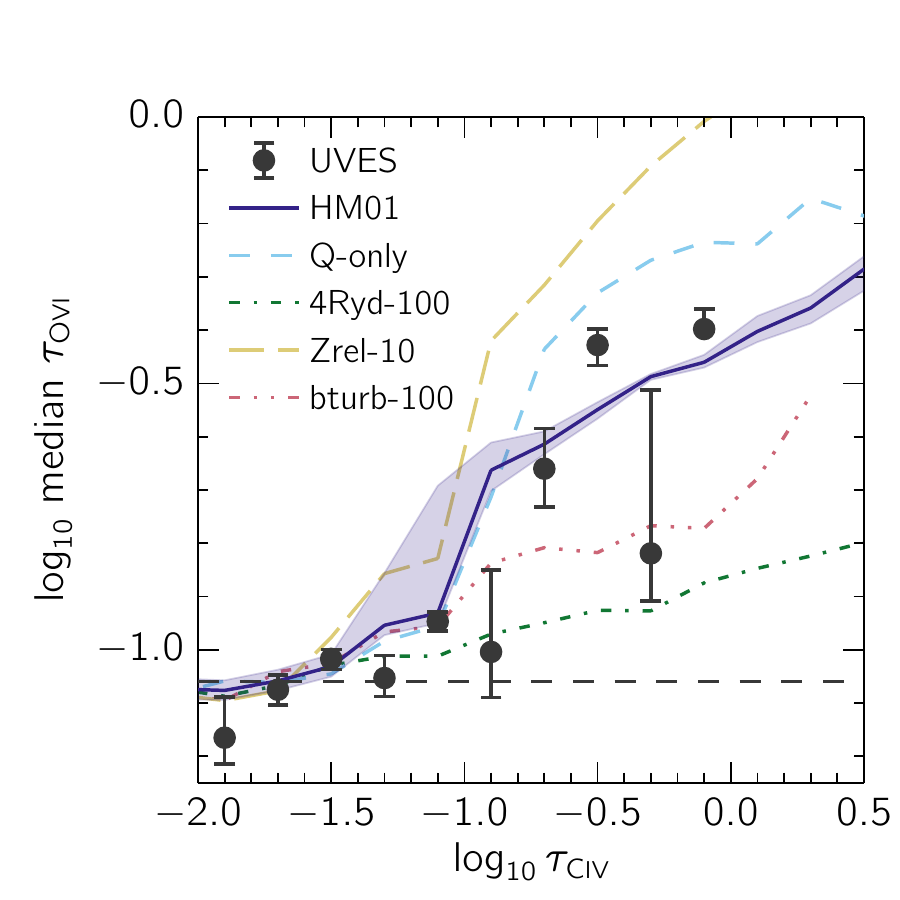}
 \includegraphics[width=\fwa]{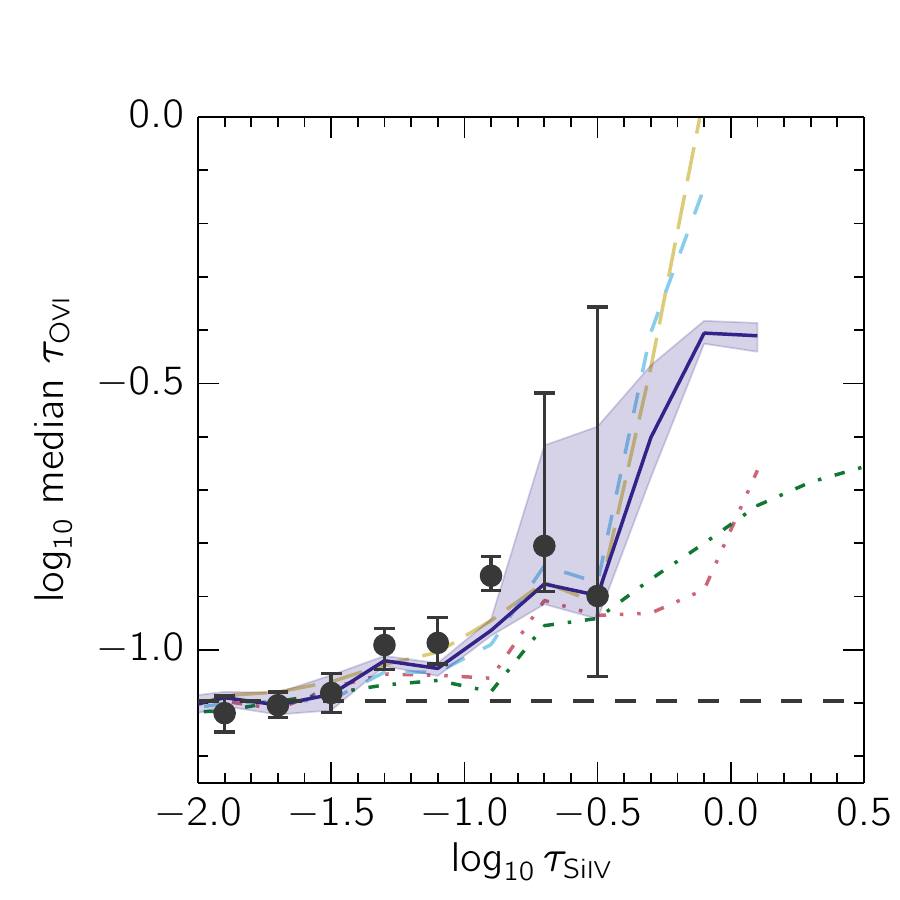}
\caption{The same as Fig.~\ref{fig:hone}, except showing \sifour(\cfour), 
   \osix(\cfour) and \osix(\sifour) from left to right, which probe relative abundances.
     The data from the observations is given in Table~\ref{tab:data_1}. 
     In the left-hand panel, we find that $\tau_{\sifourm}^{\text{med}}$ 
    is underestimated by all models except bturb-100, and is insensitive to the 
    choice of UVB.  
   Next, for  \osix(\cfour) and \osix(\sifour) we observe a stronger 
   sensitivity to different ionizing background models and turbulent broadening
   compared to \osix(\hone). For these relations, we 
   observe a better match between the fiducial and hardest UVB models (HM01 and Q-only), 
    in tension with the results from \cfour(\hone) and \sifour(\hone) relations,
    where we find a strong preference for the softer ionization backgrounds (see Fig.~\ref{fig:hone}).
     }
\label{fig:rel}
\end{figure*}

In  Fig.~\ref{fig:rel} we examine 
relations between different metal ions, which trace relative abundances
and physical conditions. The data for this figure is provided in Table~\ref{tab:data_1}. 
For example, Si/C, which can be estimated using the \sifour(\cfour) relation,
has been found to be greater than solar by a 
factor of a few in the IGM \citep[e.g.,][]{songaila01, boksenberg03, aguirre04}.

In the left-hand panel of Fig.~\ref{fig:rel}, we plot the median 
\sifour\ optical depth against \cfour. While the results are not very 
sensitive to the choice of ionizing background, all  models except bturb-100
present a paucity of \sifour\ with respect
to the observations. In particular, the median \sifour\  optical depth from the  Zrel-10 model 
shows an offset of $\leq-1$~dex from the observations at fixed
\cfour.  Again, this is due to the fact that for \sifour,
many more pixels that contribute to the median optical depth are noise dominated compared to \cfour,
and therefore do not change when the metallicity is increased.
For the remaining models that use the metallicities directly from the simulations, 
this may indicate that at $z\sim3.5$ the simulations have lower [Si/C] than observed.
Additionally, as evidenced by the bturb-100 model, turbulent 
broadening, which has the strongest influence on the heavy silicon atoms, 
could perhaps be invoked to alleviate this discrepancy.

We briefly draw attention to the bin centred at
$\log_{10}\tau_{\cfourm}=-0.3$, where the observed median \sifour\ optical
depth deviates starkly from the rest of the points. The same behaviour
is also seen in the central panel of Fig.~\ref{fig:rel}, in which
we examine \osix(\cfour). To find the origin of this inconsistency, 
we turn to the relations of individual QSOs, in Figs.~\ref{fig:app_si4_vs_c4} and
\ref{fig:app_o6_vs_c4}. In the case of Q1317$-$507 (the upper right-hand panel
of both figures), the median \sifour\ and \osix\ optical depths are 
unusually low in this \cfour\ bin, while having relatively
small error bars (the median optical depths of different QSOs are 
combined in linear space). We conclude that these points from Q1317$-$507, 
likely the result of small number statistics, are responsible for the 
anomaly in the \sifour(\hone) and \cfour(\hone) relations.

The centre panel of Fig.~\ref{fig:rel}
shows $\tau_{\osixm}^{\text{med}}$ binned by $\tau_{\cfourm}$. 
In contrast to the \osix(\honem) relation (Fig.~\ref{fig:hone}, right-hand panel), 
it is apparent that the median \osix(\cfour) optical depth
depends strongly on the choice of UVB, and is sensitive to both an increase in the elemental
abundances and the addition
of turbulent broadening. This is consistent
with the picture that \cfour\ primarily traces photoionized gas, 
which will depend on the choice ionizing background, and will not be 
as thermally broadened as hot, collisionally ionized gas. 
We find that the fiducial HM01 model is in broad agreement
with the data for this relation, even when including the bin centred
at $\log_{10}\tau_{\cfourm}=-0.3$,
while softer UVB models predict too weak \osix\ at high $\tau_{\cfourm}$.
If the \osix\ that is coincident with strong \cfour\ were photoionized, 
then this would be a useful constraint. However, unlike \cfour, \osix\ may
well be collisionally ionized.

Finally, in the right-hand panel of Fig.~\ref{fig:rel} we show 
\osix(\sifour). Except for $\log_{10}\tau_{\sifourm}\gtrsim-0.5$, we observe a much weaker 
dependence on the models than for \osix(\cfour), but we still find that 
HM01, in addition to Q-only and Zrel-10, provides the best match to the data.

\subsection{Physical conditions of the gas}

\begin{figure*}
\begin{center}
\includegraphics[width=\fwb]{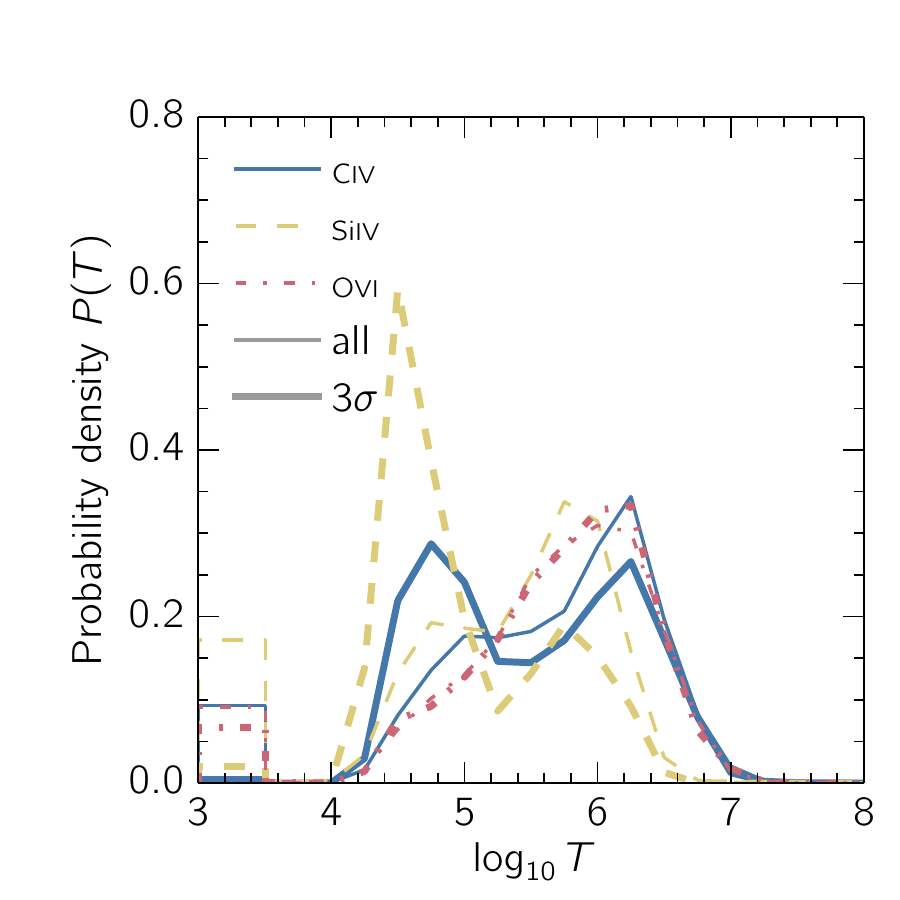}
\includegraphics[width=\fwb]{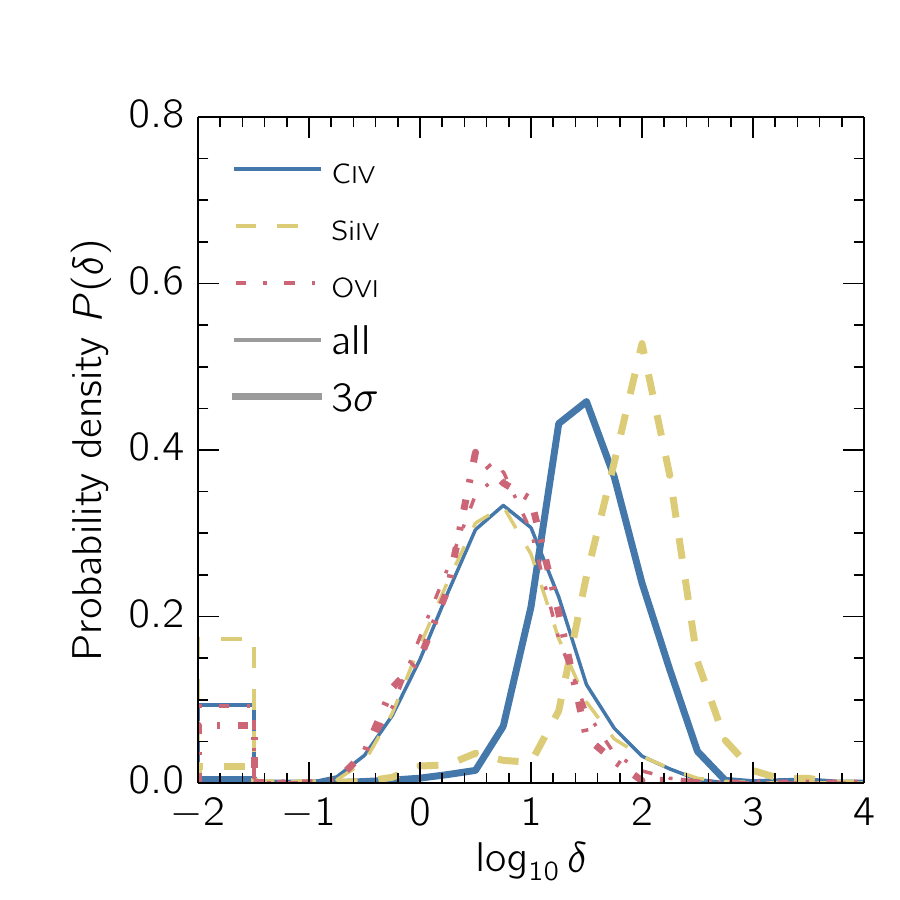}
\end{center}
 \caption{PDFs of the optical depth-weighted temperature (left) and overdensity (right), where the line
 colour and style indicates the ion used for weighting. The narrower lines represent the full 
 sample of pixels that have $\tau_{\honem}>10$, and the thicker line PDFs were
 calculated using this same sample but with the additional constraint that the metal-line optical
 depth had to be $3\sigma$ above the flat level, \taumin. This second cut is used to isolate pixels that 
 have a significant optical depth detection. The PDF offset to the left of the main distribution denotes
 pixels that have zero temperature and density. By definition these pixels have zero metallicity,
 but a detected optical depth due to noise or contamination.  We find that \cfour\ and \sifour\ probe
 a bimodal range of temperatures, while pixels with higher \osix\ optical depths
 primarily arise in hot gas. 
 }
\label{fig:pdf}
\end{figure*}

In this section, we examine the temperatures and densities of the
gas probed by our optical depth relations. In Fig.~\ref{fig:pdf}, we have plotted
the probability density functions (PDFs) of the optical depth-weighted temperatures (left) and 
densities (right) from our simulations. 
In particular, we are interested 
in the physical properties of the regions from which we detect signal in Fig.~\ref{fig:hone}.
We therefore only consider pixels with 
$\tau_{\honem}>10$, to focus on areas where 
we find the largest discrepancy between 
observations and mock spectra in this figure.\footnote{While we have used the 
recovered optical depths for the \hone\ and 
metal-line optical depth cuts made in Fig.~\ref{fig:pdf}, we note that the results 
are unchanged when we use the true optical depths instead.}

The resulting PDFs are shown as the solid lines in Fig.~\ref{fig:pdf}, and demonstrate 
 high temperatures, which may be surprising for \sifour\ and \cfour, which we
expected to be at least somewhat photoionized. 
However, these PDFs are biased to high temperatures because higher temperature gas is 
more broadened. For a single absorption line at high temperature, the optical depths will hence
be spread over more pixels, and the individual pixel optical depth values will be lower
than for a lower temperature region. 
This effect is further amplified by the fact that the pixel optical depth-weighted temperatures are averages 
over the linear (rather than log) temperatures of different gas elements.  This  means
that gas elements with relatively low ion fractions but high temperatures can affect the 
weighted means. Furthermore, at low density the ionization fraction peaks with temperature 
become much less prominent if photoionization is included (see e.g. Fig.~1 of \citealt{rahmati16}).

In an effort to combat this bias, 
we then make an additional cut, where we only take pixels that have a metal-line optical 
depths $3\sigma$ above the corresponding flat level \taumin. 
The result of making this additional cut is shown as the dashed line in Fig.~\ref{fig:pdf}. 
First, we find that the temperature and density PDF for \osix\ is unchanged, 
which indicates that most of the signal for \osix\ truly comes from gas with 
higher temperatures (although the distribution is still quite broad, consistent
with \citealt{oppenheimer16}). On the other hand,  for \cfour- and \sifour-weighted
quantities this cut reveals a bi-modal temperature distribution, where many of the pixels probe 
cooler, $\approx10^{4.5}$~K gas. 
Likewise, the \osix-weighted overdensities are not affected by the $3\sigma$ cut,
while for \cfour\ and \cfour\ the overdensity PDFs are still unimodal but have shifted to higher values. 
Overall, this figure indicates that while most of the signal for \osix(\hone) comes from
hot ($T\gtrsim10^5$) gas with $\delta\sim1$--$10$, a substantial portion of the pixels that lead to the
\cfour(\hone) and \sifour(\hone) relations arise from cooler, likely photoionized
gas at $\sim10^{4.5}$~K with overdensities $\gtrsim10$.


\section{Discussion}
\label{sec:discussion}

In the previous section, we compared observations of pixel optical depth
relations to the EAGLE simulations. We considered a fiducial QSO+galaxy HM01 UVB \citep{haardt01},
as well a harder QSO-only model, and a softer UVB with 
reduced intensity above 4~Ryd by a factor of 100.
For \osix(\hone), we found an insensitivity to the ionizing background model, and 
saw good agreement between the simulations and the data. However,
the observed median optical depths from the \cfour(\hone) and \sifour(\hone) relations
were measured to be systematically higher than those derived from the simulations
using the fiducial UVB. The discrepancy is smaller than  $\approx0.5$~dex below $\tau_{\honem}=10$ 
but can reach up to 1~dex for \hone\ bins above this threshold. For \sifour(\hone), invoking 
4Ryd-100 fully alleviates the tension, while for \cfour(\hone) 
we find this model still falls short of the data, but only for $\tau_{\honem}\gtrsim10$. 
We also find that increasing the metallicity by a factor of ten (Zrel-10)
and manually broadening the absorption lines to take unresolved turbulence into consideration (bturb-100)
do not fully resolve the discrepancy. 
In this section, we discuss in more detail possible reasons for the observed
mismatch. 

Can the discrepancies between the observations and simulations
be attributed to differences in the UVB?
We have indeed found better agreement with the observed \cfour(\hone) and \sifour(\hone)
relations using our softest UVB intended to explore the effects of delayed
\hetwo\ reionization, 4Ryd-100. The reduced 
intensity above 4~Ryd disfavours ionization to higher states, 
increasing the abundances of \sifour\ and \cfour. 
While \citet{haardt01} models take \hetwo\ reionization into account
and predict that the \hethree\ fraction already reaches 50\% at $z\approx6$, 
recent studies suggest that the reionization process is patchy,
with \hetwo\ optical depths still high above $z\gtrsim3$
\citep[e.g.,][]{shull10, worseck11}. Thus, the work presented here probes
the epoch where the observed gas may be subject to a strongly fluctuating UVB
above 4~Ryd.
The much better match of the 4Ryd-100 UVB suggests that 
\hetwo\ reionization could be complete too early in the simulations. 
Turning to other optical depth relations, we find that \cthree(\cfour)
and \sithree(\sifour) do not strongly rule out the 4Ryd-100 
model. While these soft UVBs are inconsistent with \osix(\cfour), 
the problem occurs only for $\log_{10}\tau_{\cfourm}\gtrsim-0.7$, which 
is higher than relevant for Fig.~\ref{fig:hone}. 

An alternative effect could be the presence of ionization due 
to stellar light from nearby galaxies, which is thought to be important for absorbers as rare as Lyman 
limit systems \citep{schaye06, rahmati13b}. The strength of the ionizing radiation
emitted by galaxies drops sharply above 4~Ryd, but could strongly ionize \hone,
lowering the typical optical depths. If \hone\ optical depths are lower,
than at a fixed \hone\ the metal-line optical depths will be higher.
This could explain the larger discrepancy seen at 
$\tau_{\honem}\gtrsim10$, where the pixel optical depths are probing denser gas at small
galactocentric distances compared to lower \hone\ optical 
depths. However, since it is difficult to estimate the shape 
and normalization of this ionizing radiation (and it likely should
not be applied uniformly), we leave testing of this explanation 
to a future work. 

We have also considered the effects of turbulent broadening. 
It is certainly true that our fiducial model misses the unresolved  
 turbulence from the dense particles
on the imposed EoS (where by fixing the temperatures
to $10^4$~K we neglect a possibly
significant fraction of the energy), and  possible
that turbulence in other regions is also underestimated.
Indeed, we find that by artificially broadening metal ion absorption lines, we are able to 
bring the \cfour(\hone) and \sifour(\hone)
relations into slightly better agreement with observations. Furthermore, the inclusion of $b_{\rm turb}$ 
may help alleviate the tension between the observed and simulated \sifour(\cfour).  However, we stress that our implementation
provides a very conservative upper limit on this effect, because (a) we use a very high
$b$-parameter (100~\kmps) and (b) we apply the turbulent term to all metal-line absorption pixels, 
not just those in very high density regions or with contributions from particles on the EoS. 

Another possibility for the observed discrepancy 
is that the metallicity of the intergalactic gas in the simulations
is too low. Using our Zrel-10 model, we have examined the effect of increasing
the elemental abundances linearly by a factor of ten. While such a change in metallicity is 
larger than the expected metal yield uncertainties,
it is still unable to increase the simulated median \cfour\ and \sifour\ at fixed
\hone\ enough to agree with observations. Furthermore, increasing the 
metallicities of both carbon and silicon by equal amounts leads to disagreement
between the simulated and observed \sifour(\cfour) relation, since the optical depths
do not scale directly with metallicity in the same way due to differences in 
contamination and noise. 

A related issue may be that the that the metals are not transported far enough 
into the IGM. An insufficient volume filling fraction of 
enriched gas could arise if the simulations 
do not resolve the low-mass galaxies thought to be important for metal
pollution \citep[e.g.,][]{wiersma10, booth12}. In Appendix~\ref{sec:restest}, 
we examined results from simulations with higher resolution than our fiducial model
 (the Ref- and Recal-L025N0752 runs). These simulations can resolve galaxies (containing 
at least 100 star particles) with $M_{\star}=2.3\times10^7$~\msol, 
almost an order of magnitude below that of our fiducial model, where a 
100 star particle galaxy would have stellar mass of $1.8\times10^8$~\msol. 
Indeed, we find that relations involving \cfour\ are not fully converged
at our fiducial resolution, and invoking the highest-resolution model
for \cfour(\hone) results in an increase in $\tau_{\cfourm}^{\text{med}}$ 
of up to $\approx0.3$~dex  in the highest \hone\ bins. 

To investigate the reason for the resolution dependence, we have imposed the same metallicity-density
relation on both Ref-L025N0752 and Ref-L025N0376, finding no significant differences in the optical depth relations. 
This implies that the better agreement with observations with increasing resolution is not caused by changes in the 
density-temperature structure of the gas. We have also calculated the average metallicites for the various resolutions
in the 25~cMpc volume, and find that the differences are small (varying at most by a factor of 1.2).  
Furthermore, we find that the mean gas-particle metallicity is lowest in Recal-L025N0752, and highest in Ref-L025N0188. 
Therefore, the increase in \cfour(\hone) with resolution is not due to an increase in the total amount of metals, 
but rather to an increase of the metallicity of the IGM. This suggests that winds ejected from galaxies with stellar masses below
$\approx1.8\times10^8$~\msol\ are likely important for IGM pollution. 
While a simulation with higher resolution  may bring the observations
and simulations closer to agreement, the effect does not appear to be strong enough to 
fully explain the differences seen in the \cfour(\hone) relation, and furthermore, 
the \sifour(\hone) relation shows almost no change when the resolution is
increased. Therefore, we believe that additional factors may be at play.

An important piece of information to consider is the much better
agreement between the observed and simulated \osix(\hone) relations.
The insensitivity of $\tau_{\osixm}^{\text{med}}$ (when binned by \hone) to the different UVB models 
suggests that the gas is primarily collisionally ionized, and hence that \osix(\hone)
is probing a hotter ($T\gtrsim10^5$~K) gas phase than \cfour(\hone) and \sifour(\hone),
as also found  for $\tau_{\honem}>10$ by \citet{aguirre08}. 
From this, we can conclude that for \textit{hot} gas, the physical properties probed by the 
pixel optical depth relations are consistent with observations of
the IGM at $z\sim3.5$. The lack of \cfour\ and \sifour, on the other hand,
may not be due to a too low metallicity or volume filling fraction, 
but rather to an incorrect gas phase. If too much of the enriched gas
is excessively hot, then too much carbon and silicon will be ionized to states above
 \cfour\ and \sifour, reducing the number of pixels with detectable 
 \cfour\ and \sifour\ absorption. 
 
\citet{aguirre05} found an even more severe underestimation
of simulated median \cfour\ optical depths, with the tension also
being alleviated by invoking a softer UVB. In contrast to EAGLE, 
the simulations in \citet{aguirre05} did not include metal-line cooling,
and that study found that most of the metals resided in an unrealistically-hot gas
phase ($10^5\lesssim T \lesssim 10^7$~K). The authors speculated that the simulations
could be brought into agreement with the observations by implementing
metal-line cooling, but here we have shown that this is not the case.
However, the inclusion of metal-line cooling may have aided in resolving
other issues. While we find good agreement between our observations and simulations
of the \cthree(\cfour) relation, \citet{aguirre05} measured a far too 
low $\tau_{\cthreem}^{\text{med}}$, indicating a much stronger mismatch
in the temperature and/or density of the gas in their simulations. 

It may be that the temperature of the metal-enriched gas in our 
simulations is sensitive to the 
details of the stellar feedback. 
It is implemented thermally, using a 
stochastic prescription in which the temperature of the directly heated
gas is guaranteed to initially exceed $10^{7.5}$~K \citep{dallavecchia12}. The probability
of heating events was calibrated to observations
of galaxy stellar masses and disc sizes at $z\approx0$, 
but observations of the CGM were not considered.
In Fig.~\ref{fig:pdf} we established that much of the signal for the 
\cfour(\hone) and \sifour(\hone)
relations comes from pixels with temperatures $T~\sim10^{4.5}$~K and 
overdensities $\delta\gtrsim10$. This suggests that at $z\sim3.5$, the outflows driven
by stellar feedback may not entrain enough cool ($T\sim10^4$~K) gas. 
While \citet{furlong15} found that galaxy star 
formation rate densities and stellar masses are in good agreement with observations 
for $z\sim3$--$4$, the work presented here suggests that other indicators may be  needed to
test fully the feedback implementation at these redshifts.

\section{Conclusion}
\label{sec:conclusion}

In this work we used pixel optical depth relations to study the $z\sim3.5$ IGM,
using new, very high-quality data for a sample of eight $\langle z_{\text{QSO}} \rangle=3.75$ QSOs,
and compared our results with the EAGLE hydrodynamical simulations of galaxy formation.
The QSOs were observed with VLT/UVES, and their spectra all have similar 
S/N and coverage. We employed the pixel optical depth technique to obtain
\hone\ and metal-line absorption partially corrected for the effects of 
noise, contamination, and saturation. A public version of the code used
can be found at 
\url{https://github.com/turnerm/podpy}.
 The resulting pixel optical depth
relations were compared to those derived from mock spectra
generated from the EAGLE simulations. The mock spectra were synthesized to have a
 resolution, pixel size, S/N and wavelength coverage closely matched to the observations. 
 We have considered a fiducial QSO+galaxy UVB \citep{haardt01},
as well as a harder QSO-only model and a model for which the intensity above 4~Ryd was reduced by 
a factor of 100. 
The fiducial EAGLE model was run in a cosmologically 
representative box size (100~cMpc) at a relatively high resolution ($2\times1504^3$ particles), and 
the feedback from star formation and AGN was calibrated to reproduce the 
$z\approx0$ galaxy stellar mass function, galaxy-black hole mass relation,
and galaxy disc sizes. 
Our conclusions are listed below.

\begin{itemize}
      \item We detect strong correlations for the observed median
	\cfour(\hone), \sifour(\hone), \osix(\hone) pixel optical depth relations (Fig.~\ref{fig:hone}), as well as for 
	\cthree(\cfour) and \sithree(\sifour) (Fig.~\ref{fig:temp}), and for
	\sifour(\cfour), \osix(\cfour) and \osix(\sifour) (Fig.~\ref{fig:rel}).
     \item We find that for the \cfour(\hone) and
	\sifour(\hone)  relations, the observed metal-line optical
	 depths are higher than in the simulations run with the fiducial HM01 UVB.
	 For \cfour(\hone), we find a discrepancy of up to $\approx0.1$~dex at $\tau_{\honem}=1$, 
	 $\approx0.5$~dex at $\tau_{\honem}=10$, and $\approx1$~dex at $\tau_{\honem}=10^2$, 
	 where we believe we are probing gas at high densities and small galactocentric 
	 distances. For \sifour(\hone), while the agreement is slightly better, 
	 the behaviour is qualitatively similar to that of \cfour(\hone),
	 and we find that the observed optical depths are higher than seen in the simulations
	 by up to $\approx0.2$~dex at $\tau_{\honem}=10$ 
	 and by up to $\approx0.8$~dex at $\tau_{\honem}=10^2$.
	 In contrast, \osix(\hone), which probes a hotter gas phase, exhibits
         much better agreement (i.e. differences smaller than 0.2~dex) with the data for all \hone\ bins (Fig.~\ref{fig:hone}).
     \item We consider UVBs that differ from the fiducial HM01 model, including a harder quasar-only background
	(Q-only) and softer backgrounds with 100 times reduced intensity above 4~Ryd (4Ryd-100). 
	The softer models, which may be more realistic
	than our fiducial background if \hetwo\ is still partially ionized at $z\sim3.5$,
	are a better match to the \cfour(\hone) and \sifour(\hone) relations, 
	and can nearly reproduce the observations for  $\tau_{\honem}\lesssim10$.  
	The results of the \osix(\hone) relation are however insensitive to the change    
	in UVB, which suggests that \osix\ is tracing predominantly collisionally ionized gas (Fig.~\ref{fig:hone}). 
      \item We also test a model where the elemental abundances are increased
	linearly by a factor of 10 (Zrel-10), and a model where the absorption is broadened by 
	100~\kmps\  (bturb-100). These variations are meant to explore the effects of uncertainties
	in the metal yields and of unresolved turbulence, respectively. In both 
	cases we find that the simulated \cfour(\hone) and \sifour(\hone) relations are in better (but not full)
	agreement with the observations, and stress that in we have chosen very aggressive values
	for our models such that we would expect the actual effects from uncertainties to be smaller (Fig.~\ref{fig:hone}).
     \item Examining relations that investigate different ionization states of the same element, \cthree(\cfour) and 
	\sithree(\sifour), we find good agreement between the observations and simulations 
	for the fiducial  HM01 model as well as Q-only and bturb-100. However, the observations disfavour 
	the 4Ryd-100 and Zrel-10 models (Fig.~\ref{fig:temp}). 
     \item  Most models demonstrate a mild paucity of $\tau_{\sifourm}$ in the \sifour(\cfour) relation,
	which suggests the simulations may have a slightly too low [Si/C]. The two exceptions are bturb-100, which is in good
	agreement with the observations, and Zrel-100, which demonstrates substantially too little \sifour\ at 
	fixed \cfour\ (left-hand panel of Fig.~\ref{fig:rel}).
     \item Unlike \osix(\hone), the \osix(\cfour) and \osix(\sifour) relations 
	exhibit sensitivity to the UVB for $\tau_{\cfourm}\gg10^{-1}$ and $\tau_{\sifourm}\gtrsim1$,
	and we find that \osix(\cfour) is best described by the hardest models
	(the fiducial HM01 and Q-only). The dependence on the ionizing background
	suggests that strong
	\cfour\ and \sifour\ typically probe a cooler ($T\sim10^4$~K), photoionized gas phase compared to the gas traced
	by \osix(\hone) (centre and right-hand panels of Fig.~\ref{fig:rel}). 	
    \item We use the simulations to examine the PDFs of the optical depth-weighted temperatures and densities of the pixels 
	  responsible for the high optical depth values in Fig.~\ref{fig:hone}.  We find that while 
	  \cfour, \sifour\ and \osix\ all have a component probing hot $T\gtrsim10^5$ gas, 
	  the \cfour\ and \sifour\ optical depths mainly arise from a phase of 
	  cooler ($\sim10^{4.5}$~K) gas with $\delta\gtrsim10$ (Fig.~\ref{fig:pdf}).
    \item We discuss possible reasons why \cfour\ and \sifour\ optical depths with 
       associated \hone\ are underestimated by the fiducial simulations, and we consider that perhaps
       a combination of a number of explanations are responsible:
      \begin{enumerate}
          \item Ionization by local sources, which the simulations do not account for,
          may play an important role.
          Since the strength of the radiation emitted by stars typically falls sharply above 4~Ryd, 
          this would ionize \hone\ while having a much weaker effect on the metals, which 
          would increase the median metal-line absorption for a fixed \hone\ optical 
          depth (see e.g.\ \S~4.2 and Fig.~9 in \citealt{turner15}). This explanation
          is particularly viable for $\tau_{\honem}\gtrsim10$, where we may 
          be probing small galactocentric distances. 
	\item The completion of \hetwo\ reionization in the HM01 simulations may
           occur too early, or it may be too uniform, since the observations indicate that
           it could be quite patchy around $z\sim3.5$ \citep[e.g.,][]{shull10, worseck11}.
           This explanation is supported by the better agreement between the  
           4Ryd-100 model and the \cfour(\hone) and \sifour(\hone) observations. 
           However, even the 4Ryd-100 model cannot fully explain the \cfour(\hone) observations
            for $\tau_{\honem}\gtrsim10$. 
        \item The magnitude of line-broadening, particularly in dense star-forming regions,
          could be underestimated due to unresolved 
          turbulence in the simulations. While artificially adding a large turbulent broadening
          term slightly increases the median \cfour\ and \sifour\ optical depths when binned by \hone,
          the effect is not large enough to explain the observed discrepancy. 
	\item The metallicities may not be high enough in the simulations,
	  due to uncertainties in the yields. However, even scaling the metallicities 
	  by a factor of 10 is not enough to achieve agreement in the case of \cfour(\hone). Furthermore,
	  this scaling creates tension in the \sifour(\cfour) relation, due to the fact that the median 
	  recovered optical
        \item The simulations may not resolve the low-mass galaxies
	  required to pollute the diffuse IGM. We find that the highest-resolution 
	  simulations, Ref- and Recal-L025N0752, exhibit superior agreement with the observed \cfour(\hone) relation
	  by $\approx0.3$~dex at $\tau_{\honem}\approx10^2$. While resolution likely plays a role, 
	  the magnitude of the effect does not appear large enough to explain fully the discrepancy,
	  particularly for the \sifour(\hone) relation, which we find to be almost insensitive to the 
	  resolution increase.
        \item The stellar feedback in the simulations may be driving outflows that
	  contain insufficient  cool gas ($T\sim10^4$~K). The relatively good 
	  agreement between the observed and simulated 
	  \osix(\hone) relation, which probably traces collisionally ionized gas, indicates that 
	  the simulations correctly capture
	  this hotter gas phase ($T\gtrsim10^5$~K), and that it contains enough metals.
	  However, if too much of the enriched gas is hot with respect to the observations, then more 
	  \cfour\ and \sifour\ will be ionized to higher energy levels, leading to a paucity of pixels with 
	  detected $\tau_{\cfourm}$ and $\tau_{\sifourm}$.
      \end{enumerate}
\end{itemize}

Overall, while the EAGLE simulations qualitatively reproduce all of the pixel optical depth 
correlations seen in our sample of QSOs, the mock spectra are found
to have less \cfour\ and \sifour\ at a given density than in the observations. 
This suggests that the simulations are still missing one or more important components,
which we have tested in this work:
a more rigorous treatment of \hetwo\ reionization to create a softer UVB, 
the resolution required to model turbulence that contribute to line broadening,
and/or higher metallicities and volume filling factors.
However, we do not find that any
of the above models are individually able to match the observations. 
While it is possible that the addition of enhanced photoioniziation 
of \hone\ by sources close to the absorbers may play an important role, the fact that the simulations agree with
the observed \osix(\hone) relation indicates that the fiducial model is at least able
to capture the hot gas phase correctly. 
Therefore, we believe that it is likely that the outflows created by 
energetic stellar feedback in the simulations entrain insufficient cool gas. 

\section*{Acknowledgements}

This work used the DiRAC Data Centric system at Durham University,
operated by the Institute for Computational Cosmology on behalf of the STFC DiRAC HPC 
Facility (www.dirac.ac.uk). This equipment was funded by BIS National E-infrastructure 
capital grant ST/K00042X/1, STFC capital grants ST/H008519/1 and ST/K00087X/1, STFC DiRAC 
Operations grant ST/K003267/1 and Durham University. DiRAC is part of the National 
E-Infrastructure. We also gratefully acknowledge PRACE for
awarding us access to the resource Curie based in France at Tr\`{e}s
Grand Centre de Calcul. This work was sponsored by the Dutch
National Computing Facilities Foundation (NCF) for the use of supercomputer
facilities, with financial support from the Netherlands
Organization for Scientific Research (NWO). The research was supported in part by the
European Research Council under the European Union's Seventh
Framework Programme (FP7/2007-2013)/ERC grant agreement
278594-GasAroundGalaxies and the Interuniversity Attraction
Poles Programme of the Belgian Science Policy Office [AP
P7/08 CHARM]. RAC is a Royal Society URF.


\bibliographystyle{mnras} 
\bibliography{bibliography}

\begin{thebibliography}{}
\makeatletter
\relax
\def\mn@urlcharsother{\let\do\@makeother \do\$\do\&\do\#\do\^\do\_\do\%\do\~}
\def\mn@doi{\begingroup\mn@urlcharsother \@ifnextchar [ {\mn@doi@}
  {\mn@doi@[]}}
\def\mn@doi@[#1]#2{\def\@tempa{#1}\ifx\@tempa\@empty \href
  {http://dx.doi.org/#2} {doi:#2}\else \href {http://dx.doi.org/#2} {#1}\fi
  \endgroup}
\def\mn@eprint#1#2{\mn@eprint@#1:#2::\@nil}
\def\mn@eprint@arXiv#1{\href {http://arxiv.org/abs/#1} {{\tt arXiv:#1}}}
\def\mn@eprint@dblp#1{\href {http://dblp.uni-trier.de/rec/bibtex/#1.xml}
  {dblp:#1}}
\def\mn@eprint@#1:#2:#3:#4\@nil{\def\@tempa {#1}\def\@tempb {#2}\def\@tempc
  {#3}\ifx \@tempc \@empty \let \@tempc \@tempb \let \@tempb \@tempa \fi \ifx
  \@tempb \@empty \def\@tempb {arXiv}\fi \@ifundefined
  {mn@eprint@\@tempb}{\@tempb:\@tempc}{\expandafter \expandafter \csname
  mn@eprint@\@tempb\endcsname \expandafter{\@tempc}}}

\bibitem[\protect\citeauthoryear{{Aguirre}, {Schaye}  \& {Theuns}}{{Aguirre}
  et~al.}{2002}]{aguirre02}
{Aguirre} A.,  {Schaye} J.,   {Theuns} T.,  2002, \mn@doi [\apj]
  {10.1086/341580}, \href {http://adsabs.harvard.edu/abs/2002ApJ...576....1A}
  {576, 1}

\bibitem[\protect\citeauthoryear{{Aguirre}, {Schaye}, {Kim}, {Theuns}, {Rauch}
  \& {Sargent}}{{Aguirre} et~al.}{2004}]{aguirre04}
{Aguirre} A.,  {Schaye} J.,  {Kim} T.-S.,  {Theuns} T.,  {Rauch} M.,
  {Sargent} W.~L.~W.,  2004, \mn@doi [\apj] {10.1086/380961}, \href
  {http://adsabs.harvard.edu/abs/2004ApJ...602...38A} {602, 38}

\bibitem[\protect\citeauthoryear{{Aguirre}, {Schaye}, {Hernquist}, {Kay},
  {Springel}  \& {Theuns}}{{Aguirre} et~al.}{2005}]{aguirre05}
{Aguirre} A.,  {Schaye} J.,  {Hernquist} L.,  {Kay} S.,  {Springel} V.,
  {Theuns} T.,  2005, \mn@doi [\apjl] {10.1086/428498}, \href
  {http://adsabs.harvard.edu/abs/2005ApJ...620L..13A} {620, L13}

\bibitem[\protect\citeauthoryear{{Aguirre}, {Dow-Hygelund}, {Schaye}  \&
  {Theuns}}{{Aguirre} et~al.}{2008}]{aguirre08}
{Aguirre} A.,  {Dow-Hygelund} C.,  {Schaye} J.,   {Theuns} T.,  2008, \mn@doi
  [\apj] {10.1086/592554}, \href
  {http://adsabs.harvard.edu/abs/2008ApJ...689..851A} {689, 851}

\bibitem[\protect\citeauthoryear{{Becker}, {Bolton}, {Haehnelt}  \&
  {Sargent}}{{Becker} et~al.}{2011}]{becker11}
{Becker} G.~D.,  {Bolton} J.~S.,  {Haehnelt} M.~G.,   {Sargent} W.~L.~W.,
  2011, \mn@doi [\mnras] {10.1111/j.1365-2966.2010.17507.x}, \href
  {http://adsabs.harvard.edu/abs/2011MNRAS.410.1096B} {410, 1096}

\bibitem[\protect\citeauthoryear{{Boksenberg} \& {Sargent}}{{Boksenberg} \&
  {Sargent}}{2015}]{boksenberg15}
{Boksenberg} A.,  {Sargent} W.~L.~W.,  2015, \mn@doi [\apjs]
  {10.1088/0067-0049/218/1/7}, \href
  {http://adsabs.harvard.edu/abs/2015ApJS..218....7B} {218, 7}

\bibitem[\protect\citeauthoryear{{Boksenberg}, {Sargent}  \&
  {Rauch}}{{Boksenberg} et~al.}{2003}]{boksenberg03}
{Boksenberg} A.,  {Sargent} W.~L.~W.,   {Rauch} M.,  2003, ArXiv Astrophysics
  e-prints, \href {http://adsabs.harvard.edu/abs/2003astro.ph..7557B} {}

\bibitem[\protect\citeauthoryear{{Booth} \& {Schaye}}{{Booth} \&
  {Schaye}}{2009}]{booth09}
{Booth} C.~M.,  {Schaye} J.,  2009, \mn@doi [\mnras]
  {10.1111/j.1365-2966.2009.15043.x}, \href
  {http://adsabs.harvard.edu/abs/2009MNRAS.398...53B} {398, 53}

\bibitem[\protect\citeauthoryear{{Booth}, {Schaye}, {Delgado}  \& {Dalla
  Vecchia}}{{Booth} et~al.}{2012}]{booth12}
{Booth} C.~M.,  {Schaye} J.,  {Delgado} J.~D.,   {Dalla Vecchia} C.,  2012,
  \mn@doi [\mnras] {10.1111/j.1365-2966.2011.20047.x}, \href
  {http://cdsads.u-strasbg.fr/abs/2012MNRAS.420.1053B} {420, 1053}

\bibitem[\protect\citeauthoryear{{Cowie} \& {Songaila}}{{Cowie} \&
  {Songaila}}{1998}]{cowie98}
{Cowie} L.~L.,  {Songaila} A.,  1998, \mn@doi [\nat] {10.1038/27845}, \href
  {http://adsabs.harvard.edu/abs/1998Natur.394...44C} {394, 44}

\bibitem[\protect\citeauthoryear{{Cowie}, {Songaila}, {Kim}  \& {Hu}}{{Cowie}
  et~al.}{1995}]{cowie95}
{Cowie} L.~L.,  {Songaila} A.,  {Kim} T.-S.,   {Hu} E.~M.,  1995, \mn@doi [\aj]
  {10.1086/117381}, \href {http://adsabs.harvard.edu/abs/1995AJ....109.1522C}
  {109, 1522}

\bibitem[\protect\citeauthoryear{{Crain} et~al.,}{{Crain}
  et~al.}{2015}]{crain15}
{Crain} R.~A.,  et~al., 2015, \mn@doi [\mnras] {10.1093/mnras/stv725}, \href
  {http://adsabs.harvard.edu/abs/2015MNRAS.450.1937C} {450, 1937}

\bibitem[\protect\citeauthoryear{{Crain} et~al.,}{{Crain}
  et~al.}{2016}]{crain16}
{Crain} R.~A.,  et~al., 2016, preprint, \href
  {http://adsabs.harvard.edu/abs/2016arXiv160406803C} {} (\mn@eprint {arXiv}
  {1604.06803})

\bibitem[\protect\citeauthoryear{{Cullen} \& {Dehnen}}{{Cullen} \&
  {Dehnen}}{2010}]{cullen10}
{Cullen} L.,  {Dehnen} W.,  2010, \mn@doi [\mnras]
  {10.1111/j.1365-2966.2010.17158.x}, \href
  {http://adsabs.harvard.edu/abs/2010MNRAS.408..669C} {408, 669}

\bibitem[\protect\citeauthoryear{{D'Odorico}, {Calura}, {Cristiani}  \&
  {Viel}}{{D'Odorico} et~al.}{2010}]{dodorico10}
{D'Odorico} V.,  {Calura} F.,  {Cristiani} S.,   {Viel} M.,  2010, \mn@doi
  [\mnras] {10.1111/j.1365-2966.2009.15856.x}, \href
  {http://adsabs.harvard.edu/abs/2010MNRAS.401.2715D} {401, 2715}

\bibitem[\protect\citeauthoryear{{Dalla Vecchia} \& {Schaye}}{{Dalla Vecchia}
  \& {Schaye}}{2012}]{dallavecchia12}
{Dalla Vecchia} C.,  {Schaye} J.,  2012, \mn@doi [\mnras]
  {10.1111/j.1365-2966.2012.21704.x}, \href
  {http://adsabs.harvard.edu/abs/2012MNRAS.426..140D} {426, 140}

\bibitem[\protect\citeauthoryear{{Danforth} \& {Shull}}{{Danforth} \&
  {Shull}}{2008}]{danforth08}
{Danforth} C.~W.,  {Shull} J.~M.,  2008, \mn@doi [\apj] {10.1086/587127}, \href
  {http://adsabs.harvard.edu/abs/2008ApJ...679..194D} {679, 194}

\bibitem[\protect\citeauthoryear{{Durier} \& {Dalla Vecchia}}{{Durier} \&
  {Dalla Vecchia}}{2012}]{durier12}
{Durier} F.,  {Dalla Vecchia} C.,  2012, \mn@doi [\mnras]
  {10.1111/j.1365-2966.2011.19712.x}, \href
  {http://adsabs.harvard.edu/abs/2012MNRAS.419..465D} {419, 465}

\bibitem[\protect\citeauthoryear{{Ellison}, {Songaila}, {Schaye}  \&
  {Pettini}}{{Ellison} et~al.}{2000}]{ellison00}
{Ellison} S.~L.,  {Songaila} A.,  {Schaye} J.,   {Pettini} M.,  2000, \mn@doi
  [\aj] {10.1086/301511}, \href
  {http://adsabs.harvard.edu/abs/2000AJ....120.1175E} {120, 1175}

\bibitem[\protect\citeauthoryear{{Faucher-Gigu{\`e}re}, {Lidz}, {Zaldarriaga}
  \& {Hernquist}}{{Faucher-Gigu{\`e}re} et~al.}{2009}]{fauchergiguere09}
{Faucher-Gigu{\`e}re} C.-A.,  {Lidz} A.,  {Zaldarriaga} M.,   {Hernquist} L.,
  2009, \mn@doi [\apj] {10.1088/0004-637X/703/2/1416}, \href
  {http://adsabs.harvard.edu/abs/2009ApJ...703.1416F} {703, 1416}

\bibitem[\protect\citeauthoryear{{Ferland} et~al.,}{{Ferland}
  et~al.}{2013}]{ferland13}
{Ferland} G.~J.,  et~al., 2013, \rmxaa, \href
  {http://adsabs.harvard.edu/abs/2013RMxAA..49..137F} {49, 137}

\bibitem[\protect\citeauthoryear{{Ford}, {Oppenheimer}, {Dav{\'e}}, {Katz},
  {Kollmeier}  \& {Weinberg}}{{Ford} et~al.}{2013}]{ford13}
{Ford} A.~B.,  {Oppenheimer} B.~D.,  {Dav{\'e}} R.,  {Katz} N.,  {Kollmeier}
  J.~A.,   {Weinberg} D.~H.,  2013, \mn@doi [\mnras] {10.1093/mnras/stt393},
  \href {http://adsabs.harvard.edu/abs/2013MNRAS.432...89F} {432, 89}

\bibitem[\protect\citeauthoryear{{Furlong} et~al.,}{{Furlong}
  et~al.}{2015a}]{furlong16}
{Furlong} M.,  et~al., 2015a, preprint, \href
  {http://adsabs.harvard.edu/abs/2015arXiv151005645F} {} (\mn@eprint {arXiv}
  {1510.05645})

\bibitem[\protect\citeauthoryear{{Furlong} et~al.,}{{Furlong}
  et~al.}{2015b}]{furlong15}
{Furlong} M.,  et~al., 2015b, \mn@doi [\mnras] {10.1093/mnras/stv852}, \href
  {http://adsabs.harvard.edu/abs/2015MNRAS.450.4486F} {450, 4486}

\bibitem[\protect\citeauthoryear{{Haardt} \& {Madau}}{{Haardt} \&
  {Madau}}{2001}]{haardt01}
{Haardt} F.,  {Madau} P.,  2001, in {Neumann} D.~M.,  {Tran} J.~T.~V.,  eds,
  Clusters of Galaxies and the High Redshift Universe Observed in X-rays.
  (\mn@eprint {} {astro-ph/0106018})

\bibitem[\protect\citeauthoryear{{Haardt} \& {Madau}}{{Haardt} \&
  {Madau}}{2012}]{haardt12}
{Haardt} F.,  {Madau} P.,  2012, \mn@doi [\apj] {10.1088/0004-637X/746/2/125},
  \href {http://adsabs.harvard.edu/abs/2012ApJ...746..125H} {746, 125}

\bibitem[\protect\citeauthoryear{{Haas}, {Schaye}, {Booth}, {Dalla Vecchia},
  {Springel}, {Theuns}  \& {Wiersma}}{{Haas} et~al.}{2013}]{haas13a}
{Haas} M.~R.,  {Schaye} J.,  {Booth} C.~M.,  {Dalla Vecchia} C.,  {Springel}
  V.,  {Theuns} T.,   {Wiersma} R.~P.~C.,  2013, \mn@doi [\mnras]
  {10.1093/mnras/stt1487}, \href
  {http://adsabs.harvard.edu/abs/2013MNRAS.435.2931H} {435, 2931}

\bibitem[\protect\citeauthoryear{{Hopkins}}{{Hopkins}}{2013}]{hopkins13}
{Hopkins} P.~F.,  2013, \mn@doi [\mnras] {10.1093/mnras/sts210}, \href
  {http://adsabs.harvard.edu/abs/2013MNRAS.428.2840H} {428, 2840}

\bibitem[\protect\citeauthoryear{{Kollmeier} et~al.,}{{Kollmeier}
  et~al.}{2014}]{kollmeier14}
{Kollmeier} J.~A.,  et~al., 2014, \mn@doi [\apjl]
  {10.1088/2041-8205/789/2/L32}, \href
  {http://adsabs.harvard.edu/abs/2014ApJ...789L..32K} {789, L32}

\bibitem[\protect\citeauthoryear{{Lidz}, {Faucher-Gigu{\`e}re}, {Dall'Aglio},
  {McQuinn}, {Fechner}, {Zaldarriaga}, {Hernquist}  \& {Dutta}}{{Lidz}
  et~al.}{2010}]{lidz10}
{Lidz} A.,  {Faucher-Gigu{\`e}re} C.-A.,  {Dall'Aglio} A.,  {McQuinn} M.,
  {Fechner} C.,  {Zaldarriaga} M.,  {Hernquist} L.,   {Dutta} S.,  2010,
  \mn@doi [\apj] {10.1088/0004-637X/718/1/199}, \href
  {http://adsabs.harvard.edu/abs/2010ApJ...718..199L} {718, 199}

\bibitem[\protect\citeauthoryear{{Morton}}{{Morton}}{2003}]{morton03}
{Morton} D.~C.,  2003, \mn@doi [\apjs] {10.1086/377639}, \href
  {http://adsabs.harvard.edu/abs/2003ApJS..149..205M} {149, 205}

\bibitem[\protect\citeauthoryear{{Oppenheimer} et~al.,}{{Oppenheimer}
  et~al.}{2016}]{oppenheimer16}
{Oppenheimer} B.~D.,  et~al., 2016, \mn@doi [\mnras] {10.1093/mnras/stw1066},
  \href {http://adsabs.harvard.edu/abs/2016MNRAS.tmp..845O} {}

\bibitem[\protect\citeauthoryear{{Planck Collaboration} et~al.,}{{Planck
  Collaboration} et~al.}{2014}]{planck13}
{Planck Collaboration} et~al., 2014, \mn@doi [\aap]
  {10.1051/0004-6361/201321591}, \href
  {http://adsabs.harvard.edu/abs/2014A%26A...571A..16P} {571, A16}

\bibitem[\protect\citeauthoryear{{Price}}{{Price}}{2008}]{price08}
{Price} D.~J.,  2008, \mn@doi [Journal of Computational Physics]
  {10.1016/j.jcp.2008.08.011}, \href
  {http://adsabs.harvard.edu/abs/2008JCoPh.22710040P} {227, 10040}

\bibitem[\protect\citeauthoryear{{Rahmati}, {Pawlik}, {Rai{\v c}evi\`{c}}  \&
  {Schaye}}{{Rahmati} et~al.}{2013a}]{rahmati13a}
{Rahmati} A.,  {Pawlik} A.~H.,  {Rai{\v c}evi\`{c}} M.,   {Schaye} J.,  2013a,
  \mn@doi [\mnras] {10.1093/mnras/stt066}, \href
  {http://adsabs.harvard.edu/abs/2013MNRAS.430.2427R} {430, 2427}

\bibitem[\protect\citeauthoryear{{Rahmati}, {Schaye}, {Pawlik}  \&
  {Raicevic}}{{Rahmati} et~al.}{2013b}]{rahmati13b}
{Rahmati} A.,  {Schaye} J.,  {Pawlik} A.~H.,   {Raicevic} M.,  2013b, \mn@doi
  [\mnras] {10.1093/mnras/stt324}, \href
  {http://adsabs.harvard.edu/abs/2013MNRAS.431.2261R} {431, 2261}

\bibitem[\protect\citeauthoryear{{Rahmati}, {Schaye}, {Bower}, {Crain},
  {Furlong}, {Schaller}  \& {Theuns}}{{Rahmati} et~al.}{2015}]{rahmati15}
{Rahmati} A.,  {Schaye} J.,  {Bower} R.~G.,  {Crain} R.~A.,  {Furlong} M.,
  {Schaller} M.,   {Theuns} T.,  2015, \mn@doi [\mnras]
  {10.1093/mnras/stv1414}, \href
  {http://adsabs.harvard.edu/abs/2015MNRAS.452.2034R} {452, 2034}

\bibitem[\protect\citeauthoryear{{Rahmati}, {Schaye}, {Crain}, {Oppenheimer},
  {Schaller}  \& {Theuns}}{{Rahmati} et~al.}{2016}]{rahmati16}
{Rahmati} A.,  {Schaye} J.,  {Crain} R.~A.,  {Oppenheimer} B.~D.,  {Schaller}
  M.,   {Theuns} T.,  2016, \mn@doi [\mnras] {10.1093/mnras/stw453}, \href
  {http://adsabs.harvard.edu/abs/2016MNRAS.tmp..233R} {}

\bibitem[\protect\citeauthoryear{{Rakic}, {Schaye}, {Steidel}  \&
  {Rudie}}{{Rakic} et~al.}{2012}]{rakic12}
{Rakic} O.,  {Schaye} J.,  {Steidel} C.~C.,   {Rudie} G.~C.,  2012, \mn@doi
  [\apj] {10.1088/0004-637X/751/2/94}, \href
  {http://adsabs.harvard.edu/abs/2012ApJ...751...94R} {751, 94}

\bibitem[\protect\citeauthoryear{{Rosas-Guevara} et~al.,}{{Rosas-Guevara}
  et~al.}{2015}]{rosas13}
{Rosas-Guevara} Y.~M.,  et~al., 2015, \mn@doi [\mnras] {10.1093/mnras/stv2056},
  \href {http://adsabs.harvard.edu/abs/2015MNRAS.454.1038R} {454, 1038}

\bibitem[\protect\citeauthoryear{{Rudie} et~al.,}{{Rudie}
  et~al.}{2012a}]{rudie12}
{Rudie} G.~C.,  et~al., 2012a, \mn@doi [\apj] {10.1088/0004-637X/750/1/67},
  \href {http://adsabs.harvard.edu/abs/2012ApJ...750...67R} {750, 67}

\bibitem[\protect\citeauthoryear{{Rudie}, {Steidel}  \& {Pettini}}{{Rudie}
  et~al.}{2012b}]{rudie12b}
{Rudie} G.~C.,  {Steidel} C.~C.,   {Pettini} M.,  2012b, \mn@doi [\apjl]
  {10.1088/2041-8205/757/2/L30}, \href
  {http://adsabs.harvard.edu/abs/2012ApJ...757L..30R} {757, L30}

\bibitem[\protect\citeauthoryear{{Savage}, {Kim}, {Wakker}, {Keeney}, {Shull},
  {Stocke}  \& {Green}}{{Savage} et~al.}{2014}]{savage14}
{Savage} B.~D.,  {Kim} T.-S.,  {Wakker} B.~P.,  {Keeney} B.,  {Shull} J.~M.,
  {Stocke} J.~T.,   {Green} J.~C.,  2014, \mn@doi [\apjs]
  {10.1088/0067-0049/212/1/8}, \href
  {http://adsabs.harvard.edu/abs/2014ApJS..212....8S} {212, 8}

\bibitem[\protect\citeauthoryear{{Schaller}, {Dalla Vecchia}, {Schaye},
  {Bower}, {Theuns}, {Crain}, {Furlong}  \& {McCarthy}}{{Schaller}
  et~al.}{2015}]{schaller15}
{Schaller} M.,  {Dalla Vecchia} C.,  {Schaye} J.,  {Bower} R.~G.,  {Theuns} T.,
   {Crain} R.~A.,  {Furlong} M.,   {McCarthy} I.~G.,  2015, \mn@doi [\mnras]
  {10.1093/mnras/stv2169}, \href
  {http://adsabs.harvard.edu/abs/2015MNRAS.454.2277S} {454, 2277}

\bibitem[\protect\citeauthoryear{{Schaye}}{{Schaye}}{2001}]{schaye01}
{Schaye} J.,  2001, \mn@doi [\apj] {10.1086/322421}, \href
  {http://adsabs.harvard.edu/abs/2001ApJ...559..507S} {559, 507}

\bibitem[\protect\citeauthoryear{{Schaye}}{{Schaye}}{2004}]{schaye04}
{Schaye} J.,  2004, \mn@doi [\apj] {10.1086/421232}, \href
  {http://adsabs.harvard.edu/abs/2004ApJ...609..667S} {609, 667}

\bibitem[\protect\citeauthoryear{{Schaye}}{{Schaye}}{2006}]{schaye06}
{Schaye} J.,  2006, \mn@doi [\apj] {10.1086/502792}, \href
  {http://adsabs.harvard.edu/abs/2006ApJ...643...59S} {643, 59}

\bibitem[\protect\citeauthoryear{{Schaye} \& {Dalla Vecchia}}{{Schaye} \&
  {Dalla Vecchia}}{2008}]{schaye08}
{Schaye} J.,  {Dalla Vecchia} C.,  2008, \mn@doi [\mnras]
  {10.1111/j.1365-2966.2007.12639.x}, \href
  {http://adsabs.harvard.edu/abs/2008MNRAS.383.1210S} {383, 1210}

\bibitem[\protect\citeauthoryear{{Schaye}, {Theuns}, {Rauch}, {Efstathiou}  \&
  {Sargent}}{{Schaye} et~al.}{2000a}]{schaye00b}
{Schaye} J.,  {Theuns} T.,  {Rauch} M.,  {Efstathiou} G.,   {Sargent} W.~L.~W.,
   2000a, \mn@doi [\mnras] {10.1046/j.1365-8711.2000.03815.x}, \href
  {http://adsabs.harvard.edu/abs/2000MNRAS.318..817S} {318, 817}

\bibitem[\protect\citeauthoryear{{Schaye}, {Rauch}, {Sargent}  \&
  {Kim}}{{Schaye} et~al.}{2000b}]{schaye00a}
{Schaye} J.,  {Rauch} M.,  {Sargent} W.~L.~W.,   {Kim} T.-S.,  2000b, \mn@doi
  [\apjl] {10.1086/312892}, \href
  {http://adsabs.harvard.edu/abs/2000ApJ...541L...1S} {541, L1}

\bibitem[\protect\citeauthoryear{{Schaye}, {Aguirre}, {Kim}, {Theuns}, {Rauch}
  \& {Sargent}}{{Schaye} et~al.}{2003}]{schaye03}
{Schaye} J.,  {Aguirre} A.,  {Kim} T.-S.,  {Theuns} T.,  {Rauch} M.,
  {Sargent} W.~L.~W.,  2003, \mn@doi [\apj] {10.1086/378044}, \href
  {http://adsabs.harvard.edu/abs/2003ApJ...596..768S} {596, 768}

\bibitem[\protect\citeauthoryear{{Schaye} et~al.,}{{Schaye}
  et~al.}{2015}]{schaye15}
{Schaye} J.,  et~al., 2015, \mn@doi [\mnras] {10.1093/mnras/stu2058}, \href
  {http://adsabs.harvard.edu/abs/2015MNRAS.446..521S} {446, 521}

\bibitem[\protect\citeauthoryear{{Shen}, {Madau}, {Guedes}, {Mayer},
  {Prochaska}  \& {Wadsley}}{{Shen} et~al.}{2013}]{shen13}
{Shen} S.,  {Madau} P.,  {Guedes} J.,  {Mayer} L.,  {Prochaska} J.~X.,
  {Wadsley} J.,  2013, \mn@doi [\apj] {10.1088/0004-637X/765/2/89}, \href
  {http://adsabs.harvard.edu/abs/2013ApJ...765...89S} {765, 89}

\bibitem[\protect\citeauthoryear{{Shull}, {France}, {Danforth}, {Smith}  \&
  {Tumlinson}}{{Shull} et~al.}{2010}]{shull10}
{Shull} J.~M.,  {France} K.,  {Danforth} C.~W.,  {Smith} B.,   {Tumlinson} J.,
  2010, \mn@doi [\apj] {10.1088/0004-637X/722/2/1312}, \href
  {http://adsabs.harvard.edu/abs/2010ApJ...722.1312S} {722, 1312}

\bibitem[\protect\citeauthoryear{{Simcoe}, {Sargent}  \& {Rauch}}{{Simcoe}
  et~al.}{2004}]{simcoe04}
{Simcoe} R.~A.,  {Sargent} W.~L.~W.,   {Rauch} M.,  2004, \mn@doi [\apj]
  {10.1086/382777}, \href {http://adsabs.harvard.edu/abs/2004ApJ...606...92S}
  {606, 92}

\bibitem[\protect\citeauthoryear{{Songaila}}{{Songaila}}{2001}]{songaila01}
{Songaila} A.,  2001, \mn@doi [\apjl] {10.1086/324761}, \href
  {http://adsabs.harvard.edu/abs/2001ApJ...561L.153S} {561, L153}

\bibitem[\protect\citeauthoryear{{Songaila}}{{Songaila}}{2005}]{songaila05}
{Songaila} A.,  2005, \mn@doi [\aj] {10.1086/491704}, \href
  {http://adsabs.harvard.edu/abs/2005AJ....130.1996S} {130, 1996}

\bibitem[\protect\citeauthoryear{{Springel}}{{Springel}}{2005}]{springel05a}
{Springel} V.,  2005, \mn@doi [\mnras] {10.1111/j.1365-2966.2005.09655.x},
  \href {http://adsabs.harvard.edu/abs/2005MNRAS.364.1105S} {364, 1105}

\bibitem[\protect\citeauthoryear{{Springel}, {Di Matteo}  \&
  {Hernquist}}{{Springel} et~al.}{2005}]{springel05b}
{Springel} V.,  {Di Matteo} T.,   {Hernquist} L.,  2005, \mn@doi [\mnras]
  {10.1111/j.1365-2966.2005.09238.x}, \href
  {http://adsabs.harvard.edu/abs/2005MNRAS.361..776S} {361, 776}

\bibitem[\protect\citeauthoryear{{Stinson} et~al.,}{{Stinson}
  et~al.}{2012}]{stinson12}
{Stinson} G.~S.,  et~al., 2012, \mn@doi [\mnras]
  {10.1111/j.1365-2966.2012.21522.x}, \href
  {http://adsabs.harvard.edu/abs/2012MNRAS.425.1270S} {425, 1270}

\bibitem[\protect\citeauthoryear{{Tepper-Garc{\'{\i}}a}, {Richter}, {Schaye},
  {Booth}, {Dalla Vecchia}, {Theuns}  \& {Wiersma}}{{Tepper-Garc{\'{\i}}a}
  et~al.}{2011}]{tepper-garcia11}
{Tepper-Garc{\'{\i}}a} T.,  {Richter} P.,  {Schaye} J.,  {Booth} C.~M.,  {Dalla
  Vecchia} C.,  {Theuns} T.,   {Wiersma} R.~P.~C.,  2011, \mn@doi [\mnras]
  {10.1111/j.1365-2966.2010.18123.x}, \href
  {http://adsabs.harvard.edu/abs/2011MNRAS.413..190T} {413, 190}

\bibitem[\protect\citeauthoryear{{Theuns}, {Leonard}, {Efstathiou}, {Pearce}
  \& {Thomas}}{{Theuns} et~al.}{1998}]{theuns98}
{Theuns} T.,  {Leonard} A.,  {Efstathiou} G.,  {Pearce} F.~R.,   {Thomas}
  P.~A.,  1998, \mn@doi [\mnras] {10.1046/j.1365-8711.1998.02040.x}, \href
  {http://adsabs.harvard.edu/abs/1998MNRAS.301..478T} {301, 478}

\bibitem[\protect\citeauthoryear{{Theuns}, {Viel}, {Kay}, {Schaye}, {Carswell}
  \& {Tzanavaris}}{{Theuns} et~al.}{2002}]{theuns02}
{Theuns} T.,  {Viel} M.,  {Kay} S.,  {Schaye} J.,  {Carswell} R.~F.,
  {Tzanavaris} P.,  2002, \mn@doi [\apjl] {10.1086/344521}, \href
  {http://adsabs.harvard.edu/abs/2002ApJ...578L...5T} {578, L5}

\bibitem[\protect\citeauthoryear{{Trayford} et~al.,}{{Trayford}
  et~al.}{2015}]{trayford15}
{Trayford} J.~W.,  et~al., 2015, \mn@doi [\mnras] {10.1093/mnras/stv1461},
  \href {http://adsabs.harvard.edu/abs/2015MNRAS.452.2879T} {452, 2879}

\bibitem[\protect\citeauthoryear{{Turner}, {Schaye}, {Steidel}, {Rudie}  \&
  {Strom}}{{Turner} et~al.}{2014}]{turner14}
{Turner} M.~L.,  {Schaye} J.,  {Steidel} C.~C.,  {Rudie} G.~C.,   {Strom}
  A.~L.,  2014, \mn@doi [\mnras] {10.1093/mnras/stu1801}, \href
  {http://adsabs.harvard.edu/abs/2014MNRAS.445..794T} {445, 794}

\bibitem[\protect\citeauthoryear{{Turner}, {Schaye}, {Steidel}, {Rudie}  \&
  {Strom}}{{Turner} et~al.}{2015}]{turner15}
{Turner} M.~L.,  {Schaye} J.,  {Steidel} C.~C.,  {Rudie} G.~C.,   {Strom}
  A.~L.,  2015, \mn@doi [\mnras] {10.1093/mnras/stv750}, \href
  {http://adsabs.harvard.edu/abs/2015MNRAS.450.2067T} {450, 2067}

\bibitem[\protect\citeauthoryear{{V{\'e}ron-Cetty} \&
  {V{\'e}ron}}{{V{\'e}ron-Cetty} \& {V{\'e}ron}}{2010}]{veron10}
{V{\'e}ron-Cetty} M.-P.,  {V{\'e}ron} P.,  2010, \mn@doi [\aap]
  {10.1051/0004-6361/201014188}, \href
  {http://cdsads.u-strasbg.fr/abs/2010A%26A...518A..10V} {518, A10}

\bibitem[\protect\citeauthoryear{{Wendland}}{{Wendland}}{1994}]{wendland95}
{Wendland} H.,  1994, Advances Comput.\ Math., 4, 389

\bibitem[\protect\citeauthoryear{{Wiersma}, {Schaye}  \& {Smith}}{{Wiersma}
  et~al.}{2009a}]{wiersma09a}
{Wiersma} R.~P.~C.,  {Schaye} J.,   {Smith} B.~D.,  2009a, \mn@doi [\mnras]
  {10.1111/j.1365-2966.2008.14191.x}, \href
  {http://adsabs.harvard.edu/abs/2009MNRAS.393...99W} {393, 99}

\bibitem[\protect\citeauthoryear{{Wiersma}, {Schaye}, {Theuns}, {Dalla Vecchia}
   \& {Tornatore}}{{Wiersma} et~al.}{2009b}]{wiersma09b}
{Wiersma} R.~P.~C.,  {Schaye} J.,  {Theuns} T.,  {Dalla Vecchia} C.,
  {Tornatore} L.,  2009b, \mn@doi [\mnras] {10.1111/j.1365-2966.2009.15331.x},
  \href {http://adsabs.harvard.edu/abs/2009MNRAS.399..574W} {399, 574}

\bibitem[\protect\citeauthoryear{{Wiersma}, {Schaye}, {Dalla Vecchia}, {Booth},
  {Theuns}  \& {Aguirre}}{{Wiersma} et~al.}{2010}]{wiersma10}
{Wiersma} R.~P.~C.,  {Schaye} J.,  {Dalla Vecchia} C.,  {Booth} C.~M.,
  {Theuns} T.,   {Aguirre} A.,  2010, \mn@doi [\mnras]
  {10.1111/j.1365-2966.2010.17299.x}, \href
  {http://adsabs.harvard.edu/abs/2010MNRAS.409..132W} {409, 132}

\bibitem[\protect\citeauthoryear{{Worseck} et~al.,}{{Worseck}
  et~al.}{2011}]{worseck11}
{Worseck} G.,  et~al., 2011, \mn@doi [\apjl] {10.1088/2041-8205/733/2/L24},
  \href {http://adsabs.harvard.edu/abs/2011ApJ...733L..24W} {733, L24}

\bibitem[\protect\citeauthoryear{{van de Voort}, {Schaye}, {Booth}, {Haas}  \&
  {Dalla Vecchia}}{{van de Voort} et~al.}{2011}]{vandevoort11a}
{van de Voort} F.,  {Schaye} J.,  {Booth} C.~M.,  {Haas} M.~R.,   {Dalla
  Vecchia} C.,  2011, \mn@doi [\mnras] {10.1111/j.1365-2966.2011.18565.x},
  \href {http://adsabs.harvard.edu/abs/2011MNRAS.414.2458V} {414, 2458}

\makeatother
\end{thebibliography}


%

\appendix

\section{Transitions}
\label{app:trans}

In tables~\ref{tab:trans_h1} and \ref{tab:trans_metals}, we show the rest wavelengths and oscillator strengths
of all \hone\ and metal-line transitions, respectively, used to create our mock spectra. The data 
are taken from \citet{morton03}. 

\begin{table}
\caption{Information about the \hone\ transitions used to create the mock spectra, taken from \citet{morton03}. From left to 
 right the columns denote the transition, rest wavelength $\lambda$ in angstroms, and oscillator strength $f$.}
\label{tab:trans_h1}
\begin{center}
\begin{tabular}{lrr}
\hline
Transition & $\lambda$ (\AA) & $\log_{10} f$\\
\hline
\hone  &  1215.7  &  -0.38  \\
  &  1025.7  &  -1.10  \\
  &   972.5  &  -1.54  \\
  &   949.7  &  -1.86  \\
  &   937.8  &  -2.11  \\
  &   930.7  &  -2.32  \\
  &   926.2  &  -2.50  \\
  &   923.2  &  -2.65  \\
  &   921.0  &  -2.79  \\
  &   919.4  &  -2.92  \\
  &   918.1  &  -3.04  \\
  &   917.2  &  -3.14  \\
  &   916.4  &  -3.24  \\
  &   915.8  &  -3.33  \\
  &   915.3  &  -3.41  \\
  &   914.9  &  -3.49  \\
  &   914.6  &  -3.57  \\
  &   914.3  &  -3.64  \\
  &   914.0  &  -3.71  \\
  &   913.8  &  -3.77  \\
  &   913.6  &  -3.83  \\
  &   913.5  &  -3.89  \\
  &   913.3  &  -3.94  \\
  &   913.2  &  -4.00  \\
  &   913.1  &  -4.05  \\
  &   913.0  &  -4.10  \\
  &   912.9  &  -4.15  \\
  &   912.8  &  -4.19  \\
  &   912.8  &  -4.24  \\
  &   912.7  &  -4.28  \\
  &   912.6  &  -4.32  \\
\hline\end{tabular}

\end{center}
\end{table}

\begin{table}
\caption{The same as table~\ref{tab:trans_h1}, but for metal-line transitions.}
\label{tab:trans_metals}
\begin{center}
\begin{tabular}{lrr}
\hline
Transition & $\lambda$ (\AA) & $\log_{10} f$\\
\hline
\ctwo  &  1334.5  &  -0.89  \\
  &  1036.3  &  -0.91  \\
\hline
\cthree  &   977.0  &  -0.12  \\
\hline
\cfour  &  1548.2  &  -0.72  \\
  &  1550.8  &  -1.02  \\
\hline
\fetwo  &  1144.9  &  -1.08  \\
  &  1608.5  &  -1.24  \\
  &  1063.2  &  -1.26  \\
  &  1096.9  &  -1.49  \\
  &  1260.5  &  -1.62  \\
  &  1122.0  &  -1.54  \\
  &  1081.9  &  -1.90  \\
  &  1143.2  &  -1.72  \\
  &  1125.4  &  -1.81  \\
\hline
\nfive  &  1238.8  &  -0.80  \\
  &  1242.8  &  -1.11  \\
\hline
\osix  &  1031.9  &  -0.88  \\
  &  1037.6  &  -1.18  \\
\hline
\sitwo  &  1260.4  &   0.00  \\
  &  1193.3  &  -0.30  \\
  &  1190.4  &  -0.60  \\
  &   989.9  &  -0.88  \\
  &  1526.7  &  -0.94  \\
  &  1304.4  &  -1.03  \\
  &  1020.7  &  -1.55  \\
\hline
\sithree  &  1206.5  &   0.22  \\
\hline
\sifour  &  1393.8  &  -0.29  \\
  &  1402.8  &  -0.59  \\
\hline\end{tabular}

\end{center}
\end{table}

\section{Resolution tests}
\label{sec:restest}

\begin{figure*}
\includegraphics[width=\fwa]{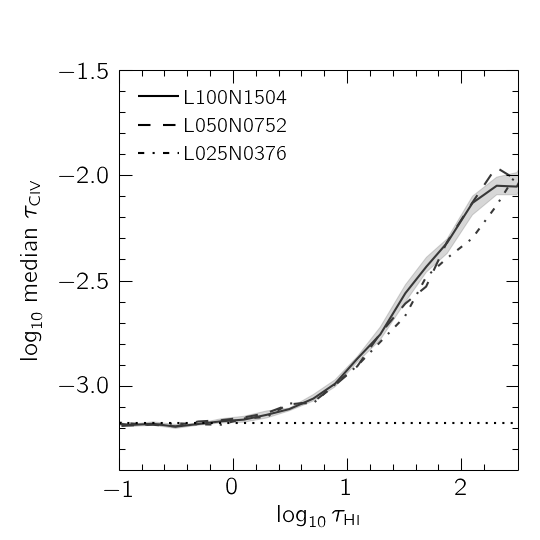}
\includegraphics[width=\fwa]{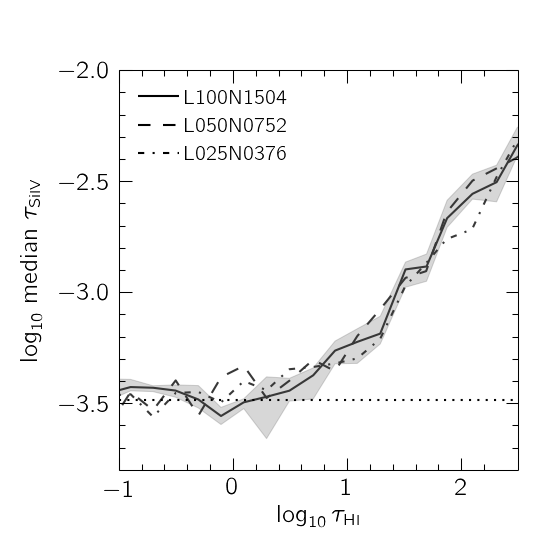}
\includegraphics[width=\fwa]{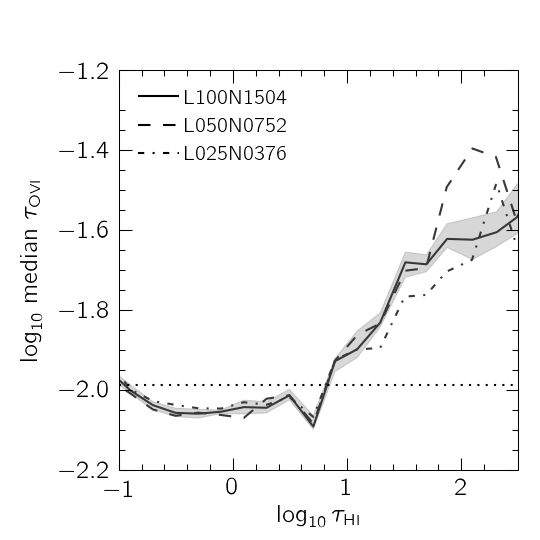}\\
\includegraphics[width=\fwa]{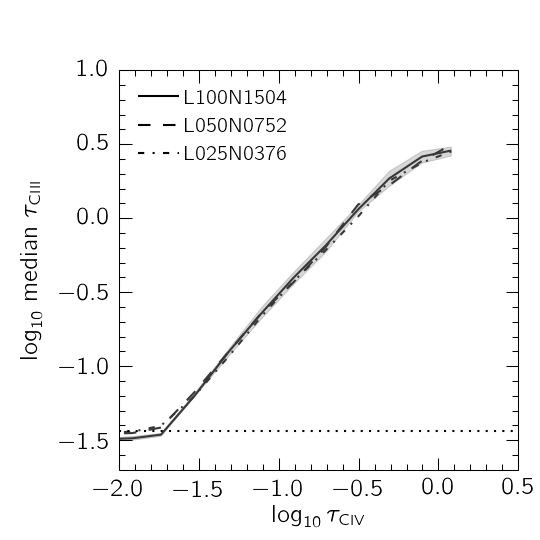}
\includegraphics[width=\fwa]{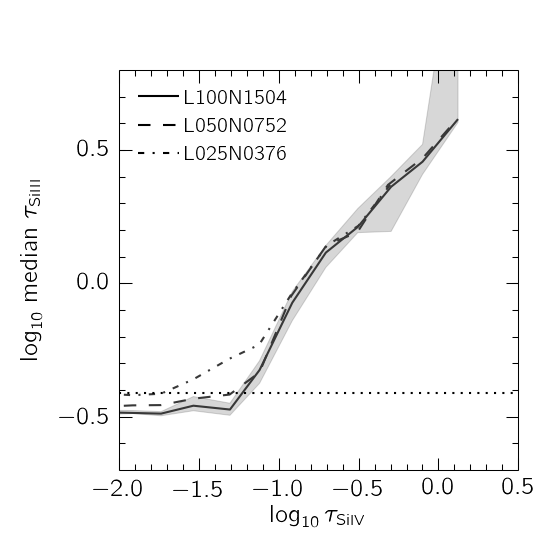}\\
\includegraphics[width=\fwa]{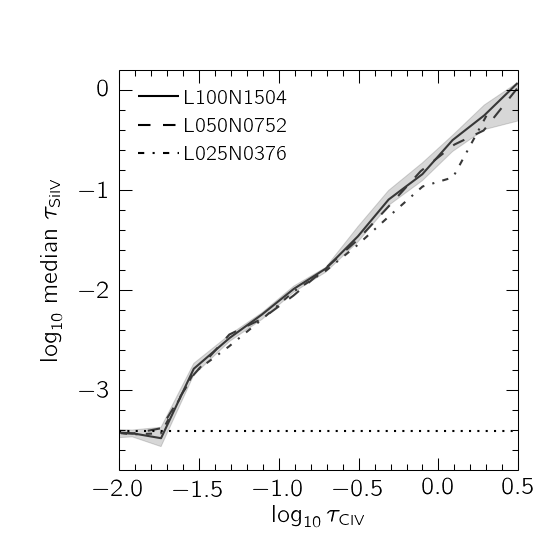}
\includegraphics[width=\fwa]{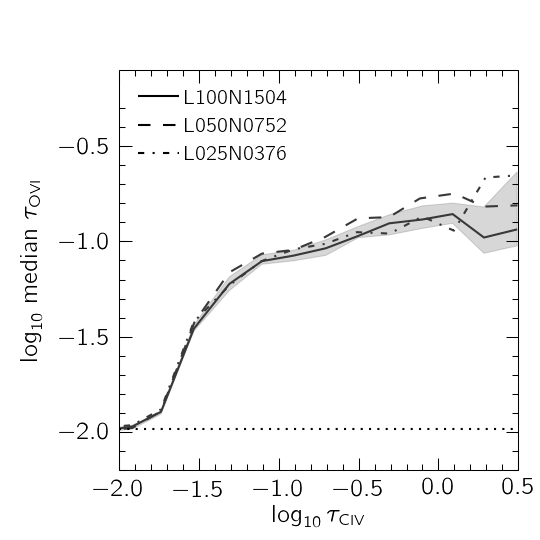}
\includegraphics[width=\fwa]{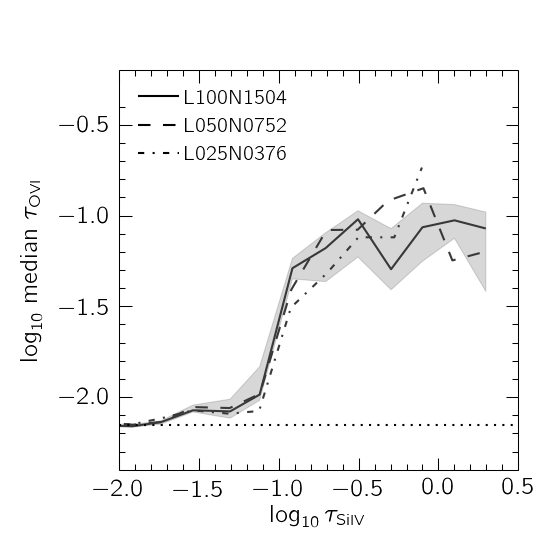}
\caption{Convergence with respect to simulation box size, where 
  we plot the same optical depth relations as were presented 
  in Figs.~\ref{fig:hone}, \ref{fig:temp}, and \ref{fig:rel}, 
  but without combining 
  different QSOs. 
  For clarity, we only show the error region around the fiducial 
  model (L100N1504), which was determined by bootstrap resampling the
  mock spectra. We find that 
    our fiducial simulation is converged.}
\label{fig:boxtest}
\end{figure*}

In this appendix, we test the numerical convergence
of the EAGLE simulations. 
We first examine the effects of varying the simulation box size.
In Fig.~\ref{fig:boxtest}, where we show optical depth relations
derived from the fiducial Ref-L100N1504 simulation,
as well as from the reference runs in 50 and 25~cMpc volumes
with the same resolution. 
To create these optical depth relations (which in this case are 
not designed to mimic observations of any particular QSO), we have generated 100 spectra
with $z_{\rm QSO}=3.94$, chosen such that the redshift of the \lya\ forest
is centred around the $z=3.53$ EAGLE snapshot. The S/N was set to be 75 throughout 
each spectrum, and the UVB was the default \citet{haardt01} model. 
We find that the optical depth relations in Fig.~\ref{fig:boxtest} are 
converged (i.e., consistent within the sample variance) for the two largest box sizes (50 and 100~cMpc).

\begin{figure*}
\includegraphics[width=\fwa]{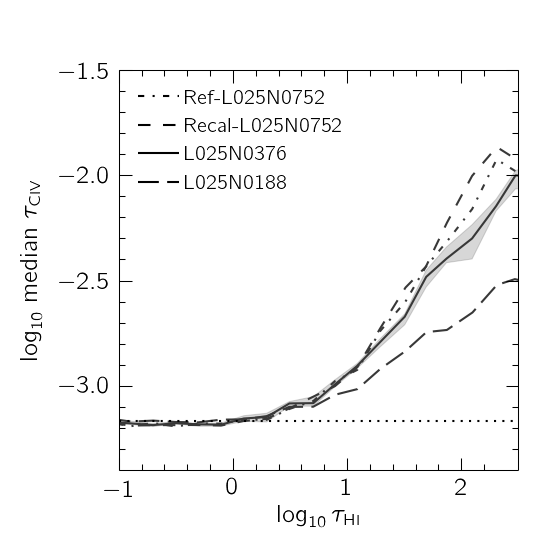}
\includegraphics[width=\fwa]{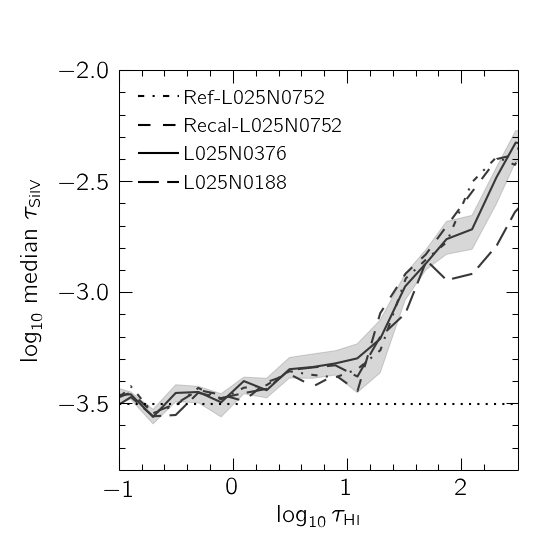}
\includegraphics[width=\fwa]{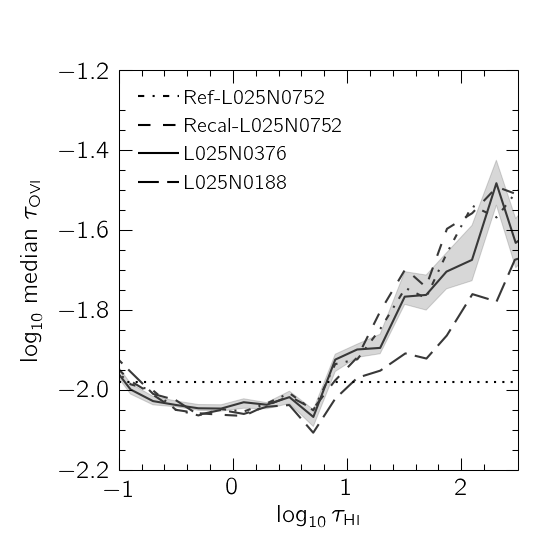}\\
\includegraphics[width=\fwa]{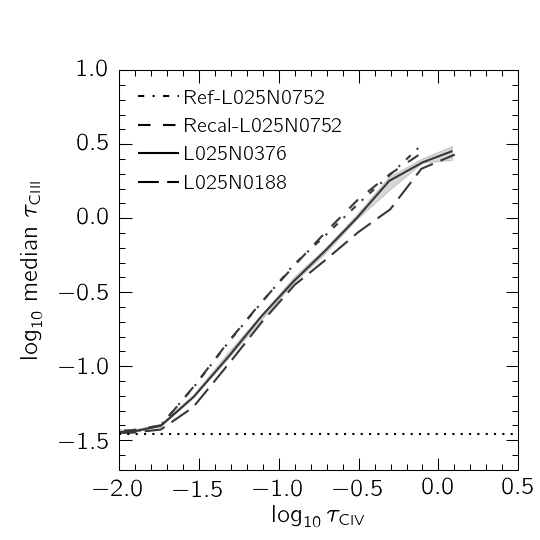}
\includegraphics[width=\fwa]{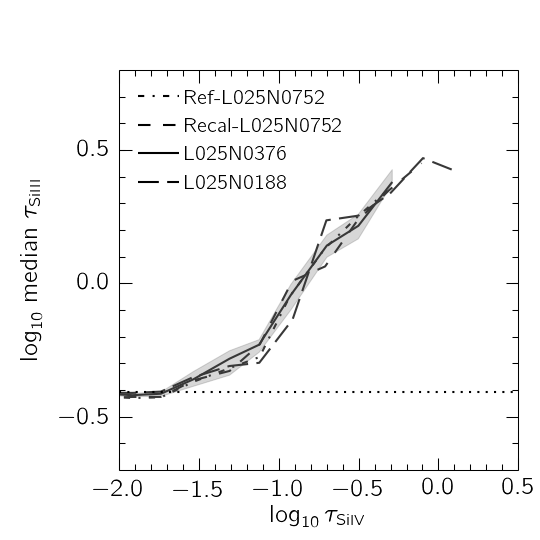}\\
\includegraphics[width=\fwa]{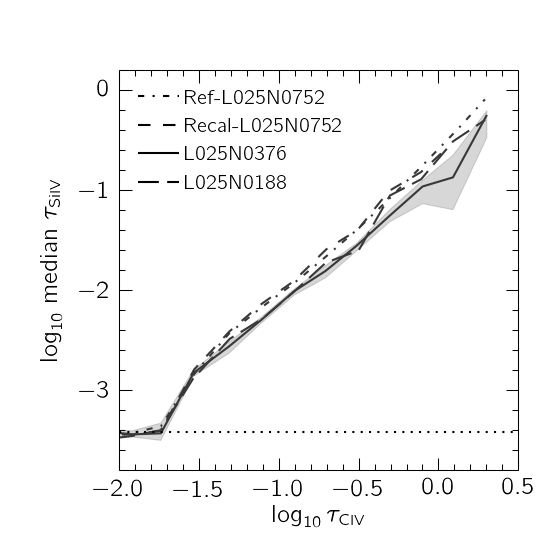}
\includegraphics[width=\fwa]{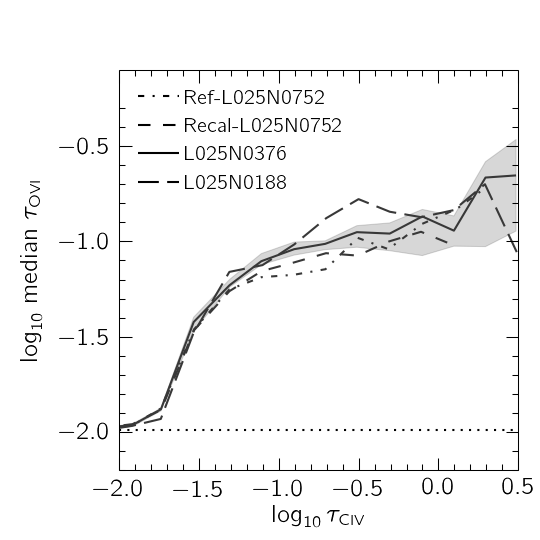}
\includegraphics[width=\fwa]{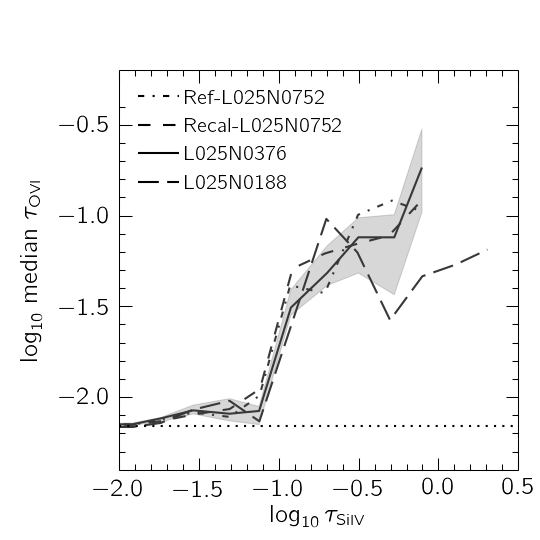}
\caption{The same as Fig.~\ref{fig:boxtest}, but showing 
 convergence with respect to the numerical resolution and using a 25~cMpc box. 
  Our fiducial resolution is given by the L025N0376 run, and indicated
  by the solid line and shaded error region.  While the
   lowest-resolution run, Ref-L025N0188, deviates significantly from the 
   others (especially for the \cfour(\hone) relation), we find mostly good agreement between 
  the fiducial intermediate- and high- resolution runs, which demonstrates
 that the fiducial resolution is nearly converged.  However, the higher-resolution runs
  predict about 0.3~dex higher $\tau_{\cfourm}^{\text{med}}$ at $\tau_{\honem}\approx10^2$, 
  which suggests that the \cfour\  and/or \hone\ associated with these rare, strong absorbers 
  have not yet converged.}
\label{fig:restest}
\end{figure*}

Next, in Fig.~\ref{fig:restest} we explore the effects of the numerical resolution, to both test for convergence
and to investigate whether pushing to lower galaxy masses may impact the enrichment of the IGM.
For this, we use the 25~cMpc box, for which simulations have been run
with resolutions higher than the fiducial one used in this work. 
There are two versions of the highest-resolution simulation L025N0752: one that has 
been run using the subgrid physics of the reference
model (Ref-) and one that has been recalibrated to better match the $z\approx0$ galaxy 
stellar mass function (Recal-). 
In the terminology of \citet{schaye15}, the comparison between Recal-L025N0752 and Ref-L025N0376 
constitutes a ``weak convergence test'' since the parameters governing the subgrid feedback 
have been recalibrated, to reproduce the z=0.1 galaxy stellar mass function. The comparison between Ref-L025N0752 
and Ref-L025N0376 constitutes a ``strong convergence test'' since the subgrid models are identical.
We include both models, as there are some interesting 
differences that may be important for the optical depth results.  In particular, the metallicity in the 
mass-metallicity relation for the Recal- model is below that of Ref-, but in better agreement
with observations \citep{schaye15}. 
We present the optical
depth relations for the above high-resolution runs, as well as for our fiducial resolution
($2\times376^3$ particles in the 25~cMpc box) and finally a lower-resolution 
of $188^3$ particles. 

In the upper left-hand panel of Fig.~\ref{fig:restest}, we examine \cfour(\hone)
and find sensitivity to resolution in the highest \hone\ bins ($\tau_{\honem}\gtrsim10^2$).
For the lowest-resolution run, the median \cfour\ optical depth
is $\approx0.5$~dex lower than for the fiducial model, while points from 
the highest-resolution simulations are up to $\approx0.3$~dex above those of
the fiducial run. For the remaining optical depths relations, the differences
are very small ($\lesssim0.1$~dex). 

These results primarily indicate that our fiducial resolution is nearly
converged, although we do find some sensitivity to 
resolution, particularly for relations involving \cfour. The suggests that a higher 
resolution results in more carbon and/or temperature conditions
that favour triply-ionized carbon. Additionally, the simulations may not be fully
converged in \hone.
However, we note that the effect of an increased resolution
is not large enough to completely resolve the discrepancy 
with observations found in Fig.~\ref{fig:hone}. Furthermore, \sifour(\hone)
shows very little sensitivity to resolution. 
  
\section{Results from single QSOs}
\label{sec:single}

In this appendix, we present the pixel optical depth relations derived
from individual QSOs, which were combined to obtain
the relations shown in Figs.~\ref{fig:hone}, \ref{fig:temp}, 
and \ref{fig:rel}. Here we display the optical depth relations
in the same order as they appear in the paper: 
\cfour(\hone) (Fig.~\ref{fig:app_c4_vs_h1}),
\sifour(\hone) (Fig.~\ref{fig:app_si4_vs_h1}),
\osix(\hone) (Fig.~\ref{fig:app_o6_vs_h1}),
\cthree(\cfour) (Fig.~\ref{fig:app_c3_vs_c4}),
\sithree(\sifour) (Fig.~\ref{fig:app_si3_vs_si4}),
\sifour(\cfour) (Fig.~\ref{fig:app_si4_vs_c4}),
\osix(\cfour) (Fig.~\ref{fig:app_o6_vs_c4}),
and 
\osix(\sifour) (Fig.~\ref{fig:app_o6_vs_si4}).

\def\reffig{c4_vs_h1}

\imgArrayRatio{c4_vs_h1}{Median \cfour\ optical depth in bins of $\tau_{\honem}$ for all eight
  QSOs. The black points represent
   the observed data, while the curves show results from simulated spectra created using different UVB models. 
   The $1\sigma$ error bars on the observations were estimated by bootstrap resampling chunks of the spectrum,
   while the error regions shown for the fiducial simulation were calculated by bootstrap resampling the 
   100 mock spectra used to generate the data. The value of \taumin\ is indicated by the horizontal dashed line for the 
   observations, and the horizontal dotted lines for the simulations.}
\imgArrayRatio{si4_vs_h1}{Same as Fig.~\ref{fig:app_\reffig}, but for \sifour(\hone).}
\imgArrayRatio{o6_vs_h1}{Same as Fig.~\ref{fig:app_\reffig}, but for \osix(\hone).}

\imgArrayRatio{c3_vs_c4}{Same as Fig.~\ref{fig:app_\reffig}, but for \cthree(\cfour).}
\imgArrayRatio{si3_vs_si4}{Same as Fig.~\ref{fig:app_\reffig}, but for \sithree(\sifour).}

\imgArrayRatio{si4_vs_c4}{Same as Fig.~\ref{fig:app_\reffig}, but for \sifour(\cfour).}
\imgArrayRatio{o6_vs_c4}{Same as Fig.~\ref{fig:app_\reffig}, but for \osix(\cfour).}
\imgArrayRatio{o6_vs_si4}{Same as Fig.~\ref{fig:app_\reffig}, but for \osix(\sifour).}

\end{document}